\documentclass[life,article,accept,moreauthors,pdftex,12pt,a4paper]{mdpi} 

\lastpage{34}
\doinum{10.3390/life4010004}
\pubvolume{4}
\pubyear{2014}
\history{Received: 1 November 2013; in revised form: 26 December 2013 / Accepted: 16 January 2014 / \\
Published: 28 January 2014}
\usepackage{amsmath}
\usepackage{amssymb}
\usepackage{url}
\usepackage{multirow}
\usepackage{graphicx}
\usepackage{xcolor}
\usepackage{placeins}
\newcommand{\degree}{\ensuremath{^\circ}}

\usepackage{soul,booktabs}

\usepackage{enumitem}
\setitemize{parsep=0pt,itemsep=0pt,leftmargin=.4in}
\setenumerate{parsep=0pt,itemsep=0pt,leftmargin=.4in}

\Title{The Formation of Jupiter, the Jovian Early Bombardment and the Delivery of Water to the Asteroid Belt: The Case of (4) Vesta}

\Author{Diego Turrini $^{1,}$* and Vladimir Svetsov $^{2}$}

\address{%
$^{1}$ Istituto di Astrofisica e Planetologia Spaziali (INAF-IAPS), Istituto Nazionale di Astrofisica, \mbox{Via del Fosso del Cavaliere 100},  Rome I-00133, Italy\\
$^{2}$ Institute for Dynamics of Geospheres, Russian Academy of Sciences, Leninskiy Prospekt 38-1,  Moscow 119334, Russia; E-Mail: svettsov07@rambler.ru}

\corres{E-Mail: diego.turrini@iaps.inaf.it; \mbox{Tel.:  +39-06-4993-4414;}
Fax: +39-06-4993-4383.}

\abstract{The asteroid (4) Vesta, parent body of the Howardite-Eucrite-Diogenite meteorites, is one of the first bodies that formed, mostly from volatile-depleted material, in the Solar System. The Dawn mission recently provided evidence that hydrated material was delivered to Vesta, possibly in a continuous way, over the last 4 Ga, while the study of the eucritic meteorites revealed a few samples that crystallized in presence of water and volatile elements. The formation of Jupiter and probably its migration occurred in the period when eucrites crystallized, and triggered a phase of bombardment that caused icy planetesimals to cross the asteroid belt. In this work, we study the flux of icy planetesimals on Vesta during the Jovian Early Bombardment and, using hydrodynamic simulations, the outcome of their collisions with the asteroid. We explore how the migration of the giant planet would affect the delivery of water and volatile materials to the asteroid and we discuss our results in the context of the geophysical and collisional evolution of Vesta. In particular, we argue that the observational data are best reproduced if the bulk of the impactors was represented by \mbox{1--2 km} wide planetesimals and if Jupiter underwent a limited (a fraction of au) displacement.}


\begin{document}

%%%%%%%%%%%%%%%%%%%%%%%%%%%%%%%%%%%%%%%%%%

\section{Introduction}

%Main text paragraph. Citing a journal paper \cite{ref-journal}. And now citing a book reference \cite{ref-book}.

The observations of the Dawn mission \cite{desanctis2012a,prettyman2012} recently confirmed the genetic link, suggested more than 40 years ago based on spectroscopic measurements \citep{mccord1970}, between the asteroid (4) Vesta and the Howardite-Eucrite-Diogenite (HED) class of meteorites. From the HED meteorites we know that Vesta was a mostly volatile-depleted body (see e.g., \cite{sarafian2013} and references therein) and, based on the crystallization ages of the oldest eucrites \citep{bizzarro2005} and diogenites \citep{schiller2011}, we know that it formed and differentiated in the first \mbox{3 Ma} of the life of the Solar System. The observational data provided by the Dawn mission \cite{desanctis2012b,denevi2012,prettyman2012} and the laboratory studies of the HED meteorites \cite{sarafian2013}, however, also revealed that water and/or hydrated materials were present on Vesta both in the most ancient and in the more recent past. The instruments on-board the Dawn spacecraft detected the presence of hydroxyls (OH) \cite{desanctis2012b} and hydrogen (H) \cite{prettyman2012}, mainly associated with the presence of dark material \cite{mccord2012,prettyman2012,desanctis2012b}, and revealed the existence of pitted terrains~\cite{denevi2012}, interpreted as the results of impact-triggered degassing events of buried volatile-rich materials. The dark material was argued to be an exogenous contaminant delivered to Vesta by the impacts of carbonaceous \mbox{chondrite-like} asteroids \cite{mccord2012,reddy2012,turrini20XX} and was suggested to be possibly linked to the existence of the pitted \mbox{terrains \cite{denevi2012}}. Alongside these evidences for the delivery of hydrated material over the last 4 Ga \mbox{(see \cite{turrini20XX}} for a discussion), HED meteorites revealed the presence of water and other volatile elements \linebreak(e.g., clorine and fluorine) in the vestan magma during the crystallization of some eucritic samples \cite{sarafian2013}. As a consequence, volatile materials should have been present, or delivered, to Vesta before the complete solidification of its eucritic crust.

%The instruments on-board the Dawn spacecraft detected the presence of hydroxyls (OH) \cite{desanctis2012b} and hydrogen (H) \cite{prettyman2012}, mainly associated to the presence of dark material \cite{mccord2012,prettyman2012,desanctis2012b}, and revealed the existence of pitted terrains \cite{denevi2012}, interpreted as the results of impact-triggered degassing events of buried volatile-rich materials. The dark material was argued to be an exogenous contaminant delivered to Vesta by the impacts of carbonaceous \mbox{chondrite-like} asteroids \cite{mccord2012,reddy2012,turrini20XX} and was suggested to be possibly linked to the existence of the pitted \mbox{terrains \cite{denevi2012}}. Alongside these evidences for the delivery of hydrated material over the last 4 Ga \mbox{(see \cite{turrini20XX}} for a discussion), HED meteorites revealed the presence of water and other volatile elements \linebreak(e.g., clorine and fluorine) in the vestan magma during the crystallization of some eucritic samples \cite{sarafian2013}. As a consequence, volatile materials should have been present, or delivered, to Vesta before the complete solidification of its eucritic crust.

According to \cite{ruzicka1997}, the thickness of the crust from which eucrites and diogenites originated should have ranged between $40$ km and $80$ km. If we follow \cite{mcsween2013} and assume an initial composition of Vesta similar to CI meteorites (olivine-rich carbonaceous chondrites named after the Ivuna meteorite), the results of \cite{ruzicka1997} would indicate that the eucritic and diogenitic layers originally had a thickness of \mbox{26 km} and 13 km, respectively. The results of thermal models (see e.g., \cite{formisano2013} and references therein for previous works on the subject by other authors) and geophysical models \cite{tkalcec2013} indicate that the eucritic crust of Vesta solidified quite early in the life of the asteroid. As discussed by \cite{formisano2013}, depending on the accretion time and the initial porosity of Vesta, the thickness of the solid crust between 3 and 5 Ma from the condensation of the CAIs (the calcium-aluminum inclusions, which are the oldest known solids in the Solar System; see \cite{coradini2011} and references therein) could have ranged from a minimum of 7--9 km to a maximum of 20--30 km. The independent study by \cite{tkalcec2013}, using a more complete physical model, gives thickness values for the solid crust varying between $7$ km and at least 10 km (\cite{tkalcec2013}, Supplementary Information; G. Golabek, personal communication) between 3 Ma and 5 Ma from CAIs, and up to 20--30 km at about 9 Ma \citep{tkalcec2013}. Therefore, the basaltic (eucritic) crust should have completely solidified somewhere between the first 3--10 Ma. In order to reach the molten layer toward the end of this temporal interval (e.g., when the solid crust was about 15 km thick) an impact would cause the formation of a $\sim$200 km wide crater, with major implications for the survival of the basaltic crust (e.g., the effusion of diogenite-rich magma and the formation of diogenitic terrains units, \cite{turrini2011,turrini2013}) that have not been observed by the Dawn mission \cite{desanctis2012a,prettyman2012}. As a consequence, whatever mechanism delivered the volatile elements incorporated in the eucritic samples studied by \cite{sarafian2013}, it should have acted during the first few Ma of the life of Vesta and it should have preserved the basaltic crust of the asteroid, which we know survived until present time \cite{desanctis2012a,prettyman2012}.

%At the time Vesta was forming its basaltic crust, the Solar System was in the phase of its evolution known as the Solar Nebula \cite{coradini2011}, \textit{i.e.}, it was a circumsolar disk of gas and dust where the first generations of planetary bodies were forming. The beginning of the Solar Nebula is generally assumed to coincide with the condensation of the CAIs about $4.6$ Ga ago (the oldest CAI currently known dates \mbox{${4568.2}^{+0.2}_{-0.4}$ Ma} ago \cite{bouvier2010}), while its duration is indirectly constrained by observation
At the time Vesta was forming its basaltic crust, the Solar System was in the phase of its evolution known as the Solar Nebula \cite{coradini2011}, \textit{i.e.}, it was a circumsolar disk of gas and dust where the first generations of planetary bodies were forming. The beginning of the Solar Nebula is generally assumed to coincide with the condensation of the CAIs about $4.6$ Ga ago (the oldest CAI currently known dates \mbox{${4568.2}^{+0.2}_{-0.4}$ Ma} ago \cite{bouvier2010}), while its duration is indirectly constrained by observations and theoretical studies of circumstellar disks, whose median (inner disk) lifetime is about $3$ Ma with the range of observed values spanning between $1$--$10$ Ma \cite{meyer2008}. The giant planets of the Solar System should have formed across this timespan (see \cite{dangelo2011} for an overview on the processes and the timescales governing the formation of giant planets) and, in particular, theoretical \citep{bottke2005a,bottke2005b} and observational \cite{scott2006} arguments suggest that Jupiter formed $3$--$5$ Ma after the condensation of CAIs. The formation of Jupiter has been shown by different authors \cite{safronov1969,weidenschilling1975,weidenschilling2001,turrini2011,turrini2012,turrini2013} to trigger a sudden spike in the flux of impactors in the early history of the Solar System. This event, named the Jovian Early Bombardment (\cite{turrini2011,turrini2012}, JEB in the following), is caused by the scattering of ice-rich planetesimals from the outer Solar System due to the gravitational perturbation of the giant planet \cite{safronov1969,weidenschilling1975,weidenschilling2001,turrini2011,turrini2012} and by the appearance of the Jovian \linebreak mean motion resonances in the asteroid belt, in particular the $3$:$1$ and $2$:$1$ resonances \cite{weidenschilling2001,turrini2011,turrini2012,turrini2013}. \linebreak The duration of the JEB is limited to about $1$ Ma \cite{weidenschilling1975,turrini2011,turrini2012}, with the bulk of the impacts taking place in the first ($3$--$5)\times10^{5}$ years \cite{turrini2011}. The flux of impactors due to the Jovian resonances is the dominant one in the inner Solar System \cite{turrini2011,turrini2013} and is the one shaping the early collisional evolution of the asteroid \mbox{belt \cite{turrini2011,turrini2012,turrini2013}.}

Turrini \textit{et al.} \cite{turrini2011} estimated the fluxes of impactors coming from the outer Solar System and from the Jovian resonances during the JEB, the crater populations they produce and the probability of Vesta being destroyed during the bombardment using different size-frequency distributions (SFDs in the following) of the impactors. Their results showed that the probability of Vesta undergoing a catastrophic \linebreak impact are negligible, but suggested that cratering erosion could play an unexpectedly significant role due to the higher, pre-depletion population of planetesimals inhabiting the asteroid belt at the time. Turrini \textit{et al.} \cite{turrini2012} further investigated the subject of asteroidal erosion during the JEB and, using a more detailed physical description of the mass loss processes, showed that cratering erosion indeed played a much more relevant role than catastrophic disruption in determining the fate of primordial asteroids. Turrini \textit{et al.} \cite{turrini2012} showed that cratering erosion is a function of the extent of Jupiter's migration and of the position of the target body in the asteroid belt. Depending on the considered scenario and SFD of the impactors, planetesimals the size of Vesta could lose from a few times $1\%$ to a few times $10\%$ of their original mass \cite{turrini2012}.  Turrini~\cite{turrini2013} re-evaluated the collisional evolution of Vesta due to the impactors from the inner Solar System by reprocessing the results of the simulations from \cite{turrini2011} with an improved version of the collisional model detailed by \cite{turrini2012} and by investigating the roles of crater saturation, surface excavation and cratering erosion. According to the results of \cite{turrini2013}, the survival of the basaltic crust of Vesta would favor scenarios where Jupiter underwent a limited (no more than 0.25 au) migration and the protoplanetary disk was dominated, in terms of population, by planetesimals with diameters of 1--2 km while most of the mass of the disk was in the form of 1000 km wide or larger planetary embryos. However, Turrini \cite{turrini2013} also pointed out that, if an undifferentiated crust survived on top of the molten interior of Vesta after the differentiation process as suggested by the results of \cite{formisano2013,tkalcec2013}, the favored scenarios would be those where migration was more significant (0.50--1.00 au) to allow for the removal of said undifferentiated crust.

The aim of this work is to evaluate the implications of the fluxes of impactors originating from the outer Solar System in the simulations of \cite{turrini2011} for the surface evolution of Vesta and the delivery of volatile materials to the asteroid. To achieve this goal, we will proceed in a way similar to what has been done by \cite{turrini2013} for the impactors originating from the inner Solar System. We will use the hydrodynamic method developed by \cite{svetsov2011} to assess the erosion caused by the impacts of icy planetesimals at different velocities and, at the same time, to evaluate the fraction of the mass of these icy impactors that would be retained by Vesta. Using these data, we will adapt the collisional model developed by \cite{turrini2013} to the case of icy impactors, we will evaluate their effects in terms of crater saturation, surface erosion and excavation for Vesta, and we will assess whether the JEB could be the source of the water and volatile elements observed in the eucritic samples studied by \cite{sarafian2013}. Finally, we will discuss the results obtained in this study together with those obtained by \cite{turrini2013} to constrain the migration of Jupiter and the size-frequency distribution of the planetesimals in the Solar Nebula.

%using the collisional model developed by \cite{turrini2013} together with a 

%%%%%%%%%%%%%%%%%%%%%%%%%%%%%%%%%%%%%%%%%%

\section{The Jovian Early Bombardment and the Collisional Model}
\label{model-jeb}

The main input data we used in this study of the delivery of water to Vesta across the JEB are the fluxes of impactors described in \cite{turrini2011}, hereafter Paper I. To assess the effects of the JEB in terms of the collisional evolution of Vesta (e.g., cratering erosion, crater saturation, excavation of the crust) we based on the collisional model described in \cite{turrini2012}, hereafter Paper II, and \cite{turrini2013}, hereafter Paper III.  \linebreak {As the dynamical}, physical and numerical details of the model used to study the JEB have been extensively described in Papers I, II and III, here we will discuss the main aspects of the simulations and highlight the differences with respect to the previous works. The interested readers are referred to Paper I for more details on the fluxes of impactors on Vesta during the JEB and to Paper III for an extensive discussion of the effects of the impactors originating from the inner Solar System on the collisional and geophysical evolution of the asteroid. The numerical model used to assess the amount of cometary material surviving each impact and the associated erosion of the vestan surface is based on the one described in \cite{svetsov2011}: in the following, the improvements and the modification to the original model will be described in more detail.

\subsection{The Solar Nebula}\label{model-nebula}

The template of the Solar System at the beginning of the simulations of Paper I was composed of the Sun, the forming Jupiter, Vesta, Ceres and a disk of planetesimals modeled as massless particles. \mbox{The evolution of} this template of the Solar System was followed for $2\times10^{6}$ years, centered on the time Jupiter started to accrete the nebular gas.%reached its final mass.

\subsubsection{The Protoplanetary Disk}%\label{model-disk}

In the simulations of Paper I the disk of planetesimals extended between $2$ au and $10$ au. %in Paper I %and between $2$ and $8$ au in Paper II. In both cases, the disk 
The disk was divided into annular regions of width equal to $1$ au, each containing $10^{4}$ massless particles. The planetesimals had initial values of the orbital eccentricity and of the inclination ranging between \linebreak $0 \leq e_{i} \leq 3\times10^{-2}$ and $0$ rad $\leq i_{i} \leq 3\times10^{-2}$ rad, respectively. Planetesimals formed in the inner Solar System (2--4 au, ISS in the following) were considered rocky bodies with mean density \mbox{$\rho_{iss}=3.0$ g cm$^{-3}$}, while planetesimals formed in the outer Solar System (4--10 au, OSS in the following) were considered volatile-rich bodies with mean density $\rho_{oss}=1.0$ g cm$^{-3}$. In Paper I the transition between ISS and OSS impactors, assumed to coincide with the location of the Snow Line, took place at $r_{SL}=4.0$ au or at $r_{SL}=3.0$ au depending on the assumed size-frequency distribution of the impactors. In this work, the transition between ISS and OSS impactors will always take place at $r_{SL}=4.0$ au. The planetesimals were removed from the simulations of Paper I if their semimajor axes became smaller than $1$ au, or larger than $30$ au, or if they impacted the Sun or Jupiter. Each massless particle could impact only once with each of the target bodies considered in Paper I (\textit{i.e.}, Vesta and Ceres) % and II 
 but was not removed from the simulations unless it impacted both targets or one of the previously listed conditions occurred.

\subsubsection{Vesta}%, Ceres and synthetic target bodies}\label{model-vesta}

In the simulations of Paper I the target bodies considered were Vesta and Ceres. The semimajor axis of Vesta was assumed $a_{v}=2.362$ au and the asteroid was initially on a planar, circular orbit. Its mean radius was assumed $r_{v}=258$ km based on \cite{thomas1997} and its mass value $m_{v}=2.70\times10^{23}$ g was derived \mbox{from \cite{michalak2000}}. Note that the values of the mean radius and of the mass adopted for Vesta in Paper I are slightly different from the ones recently estimated by the Dawn mission ($262.7$ km and $2.59\times10^{23}$ g \mbox{respectively, \cite{russell2012}}). However, these values were only used to evaluate the collisional cross-section of Vesta, and the differences between the two sets of values are of the order of only $2\%$--$4\%$, so they do not significantly affect the fluxes estimated in Paper I.

\subsubsection{Jupiter}%\label{model-jupiter}

At the beginning of the simulations, Jupiter was a planetary embryo with mass $M_{0}=0.1\,M_\oplus$ (where $M_\oplus=5.9726\times10^{27}$ g is the Earth's mass) that grew to the critical mass $M_{c}=15\,M_\oplus$ in $\tau_{c}=10^{6}$ years (where $\tau_{c}$ can be interpreted as the oligarchic growth timescale, \cite{dangelo2011}) as:
\begin{equation}
 M=M_{0}+\left( \frac{e}{e-1}\right)\left(M_{c}-M_{0}\right)\times\left( 1-e^{-t/\tau_{c}} \right)
\end{equation}

As shown in Papers I and II, the effects of Jupiter on the cometary flux on Vesta and the inner asteroid belt during this phase are limited. A few OSS planetesimals can have close encounters with Jupiter, be injected on orbits crossing the asteroid belt and impact on Vesta or one of the largest asteroids, but they are not considered in this analysis as their contribution to the total flux  on Vesta is smaller than that of OSS planetesimals during the JEB (see Paper I). As shown in Paper I, however, this approximation would not be correct in the case of Ceres and, more generally, a larger statistics (in terms of simulations) is needed in order to properly assess their role in the flux of volatile-rich impactors to the asteroid belt.

When the critical mass value $M_{c}$ is reached, the nebular gas surrounding Jupiter is assumed to be rapidly accreted by the planet, whose mass then grows as:
\begin{equation}
 M=M_{c}+\left( M_{J} - M_{c}\right)\times\left( 1-e^{-(t-\tau_{c})/\tau_{g}}\right)
\end{equation}
where $M_{J}=1.8986\times10^{30}\,g=317.83\,M_{\oplus}$ is the final mass of the giant planet. The e-folding time $\tau_{g}=5\times10^3$ years is derived from the hydrodynamical simulations described in \cite{lissauer2009,coradini2010}. While accreting the nebular gas, Jupiter migrates inward due to disk-planet interactions as:
\begin{equation}
 r_{p}=r_{0}+\left( r_{J} - r_{0}\right)\times\left( 1-e^{-(t-\tau_{c})/\tau_{r}}\right)
\end{equation}
where $r_{p}$ is the instantaneous orbital radius of Jupiter, $r_{0}$ is its orbital radius at the beginning of the simulation, $r_{J}$ is its present orbital radius and $\tau_{r}=\tau_{g}=5\times10^{3}$ years. Paper I considered four different migration scenarios: no displacement and $0.25$ au, $0.50$ au and $1.00$ au displacements. The initial position of Jupiter is chosen in the different scenarios so that its final position is always the present one, consistently with the original Nice model (\cite{tsiganis2005}, note that in the most recent developments of the Nice model \citep{levison2011} Jupiter is assumed to be initially located at about $5.4$ au). With the chosen values of the e-folding time $\tau_{r}$, the migration scenarios are equivalent to assume values of $a/\dot{a}$ (\textit{i.e.}, the radial migration timescale, see \cite{papaloizou2007}) respectively equal to $3.2\times10^{5}$ years, $1.6\times10^{5}$ years and $8\times10^{4}$ years, consistently with the results of theoretical studies \cite{papaloizou2007}. As a test, Paper I modified the values of $\tau_{g}$ and $\tau_{r}$ to $2.5\times10^{4}$ years in the scenario were Jupiter migrates by $1$ au. Neither the fluxes of impactors nor the impact velocities changed in any significant way on Vesta. The same was not true for Ceres, which received a flux of OSS impactors a factor of $4$ higher.

\subsection{Size-Frequency Distributions of Planetesimals}\label{model-sfds}

In this study we will consider a total of $4$ SFDs of the primordial planetesimals populating the Solar Nebula, whose selection is based on the results of Papers II and III, and on the criterion of the survival of Vesta and its basaltic crust \cite{davis1985}. The first two SFDs considered \cite{coradini1981,chambers2010} supply the post-formation, non-collisionally evolved average size of the primordial planetesimals as a function of their distance from the Sun. As a consequence, these SFDs can be be applied directly to both ISS and OSS impactors. The last two SFDs \cite{morbidelli2009,weidenschilling2011} provide instead the collisionally evolved size distribution of the planetesimals after $3$ Ma, assumed to be the formation time of Jupiter, but were derived exclusively for the orbital region of the asteroid belt. In principle, therefore, they cannot be applied directly to the OSS impactors that are the focus of this work.

The results of \cite{weidenschilling2011}, however, showed that for diameters down to $1$ km the collisionally evolved SFD of the primordial planetesimals varies little with distance in the orbital region between $1.5$ au and \mbox{$4$ au} (see \mbox{{Figure 14}} from \cite{weidenschilling2011}). According to \cite{weidenschilling2011}, the collisionally evolved SFD also shows little sensitivity to the assumed planetesimal density (see \mbox{Figure 13} from \cite{weidenschilling2011}) and to the surface density of the swarm (see \mbox{{Figure 19}} from \cite{weidenschilling2011}). Finally, Weidenschilling~\cite{weidenschilling2008} showed that, for diameters smaller than $100$ km, the SFD of planetesimals that underwent between 1 Ma and 5 Ma of collisional evolution is essentially similar between $0.5$ and $10$ au \linebreak (see \mbox{{Figure 4}} from \cite{weidenschilling2008}). It must be noted, however, that \cite{weidenschilling2008} considered a initial size of the planetesimals larger than that of \cite{weidenschilling2011}, \textit{i.e.}, $1$ km \textit{versus} $100$--$500$ m respectively.

Because of the previous results, the SFDs by \cite{morbidelli2009,weidenschilling2011} were included in this work to investigate the cometary bombardment in the asteroid belt in a collisionally evolved disk of planetesimals. This is equivalent to assuming that the invariance of the SFD with distance, density of the planetesimals and surface density of the swarm observed by \cite{weidenschilling2011} between $1.5$ and $4$ au also holds between $4$ au and \mbox{$10$ au}. Based on the results of \cite{weidenschilling2008} this assumption appears to be reasonable, especially for the plausible temporal range of the formation of Jupiter (3--5 Ma) and size range of the planetesimals (1--10 km) that constitutes the bulk of the JEB based on the results of Paper III. As this assumption may nevertheless be not correct, the results that will be obtained using these SFDs should be regarded as an indication and should be considered with caution. 

From the results of \cite{weidenschilling2008}, however, we can extrapolate that the main effect (for the JEB) of the uncertainties on the SFD of the OSS planetesimals would be on the relative abundances of the impactors between 1 km and 10 km in diameters. This effect would be larger at about 1 Ma after the beginning of the planetary formation process, but would become significantly less important by about 5 Ma \cite{weidenschilling2008}. Depending on the actual value of the slope of the differential size distribution of the impactors \cite{davis1979}, the uncertainty on the slope can result in either an increase or a decrease in the OSS mass flux on Vesta. Based on the results of Paper III when considering the SFDs from \cite{coradini1981,morbidelli2009,weidenschilling2011} for what it concerns the magnitude of the effects linked to the mass flux and the size of the impactors (e.g., the surface erosion), we can expect this uncertainty to change only quantitatively (by a factor of a few) but not qualitatively the results of this work.

\subsubsection{Planetesimals Formed in a Quiescent Disk}
The first SFD considered is that of a disk of planetesimals formed by gravitational instability of the dust in the mid-plane of a non turbulent protoplanetary nebula \cite{safronov1969,goldreich1973,weidenschilling1980}. The protoplanetary nebula is assumed to have a mass $M_{neb}=0.02$ M$_\odot$ (where $M_\odot=1.9891\times10^{33}$ g is the solar mass) distributed between $1$--$40$ au, dust-to-gas ratio $\xi=0.01$ inside the Snow Line and $\xi=0.02$ beyond the Snow Line, and density profile $\sigma=\sigma_{0}\left(\frac{r}{1\,au}\right)^{-n_{s}}$, where $\sigma_{0}=2700  $g cm$^{-2}$ is the surface density at $1$ au, $r$ is the heliocentric distance expressed in cm, the symbol $1$ au indicates the value of the astronomical unit expressed in cm, and $n_{s}=1.5$. The initial mass of solids contained in the region comprised between \mbox{$2$ au and $3$ au} ({\it i.e.}, the reference region considered also by \cite{morbidelli2009,weidenschilling2011}) is about $4$ M$_{\oplus}$. For such a \mbox{nebula \cite{coradini1981}} showed that the average mass of the planetesimals $\overline{m}_{p}$ would follow the semi-empirical relationship
\begin{equation}\label{masslaw}
 \overline{m}_{p}=m_{0}\left( \frac{r}{1\,au} \right)^{\beta}
\end{equation}
where $\overline{m}_{p}$ and $m_{0}$ are expressed in g, and $\beta=1.68$. The value $m_{0}$ is the average mass of planetesimals at $1$ au, \textit{i.e.}, $2\times10^{17}$ g \cite{coradini1981}. Paper I showed that, assuming that the mass dispersion of the planetesimals about the average values of Equation \eqref{masslaw} is governed by a Maxwell-Boltzmann distribution, a mass value can be associated to each test particle by means of a Monte Carlo method where the uniform random variable $Y$ varying in the range $[0,1]$ is 
\begin{equation}
 Y=\frac{2\gamma\left(3/2,y^{*}\right)}{\sqrt{\pi}}=P\left(3/2,y^{*}\right)
\end{equation}
where $P\left(3/2,y^{*}\right)$ is the lower incomplete Gamma ratio.
The inverse of the lower incomplete Gamma ratio can be computed numerically and, by substituting $y^{*}$ back with $m^{*}/\overline{m}_{p}(r)$ we obtain
\begin{equation}\label{massval}
 m(r)=\overline{m}_{p}\,inv\left(P\left(3/2,Y\right)\right)
\end{equation}
Since the use of massless particles assures the linearity of the processes investigated over the number of considered bodies, the number of impacts expected in such a disk of planetesimals is extrapolated by multiplying the number of impacts recorded in the simulations of Paper I by a factor $\gamma$ where
\begin{equation}\label{ratio}
 \gamma=N_{tot}/n_{mp}
\end{equation}
where $n_{mp}=8\times10^{4}$ and $N_{tot}$ is given by
\begin{align}\label{ntot}
 N_{tot}&=\int^{r_{max}}_{r_{min}} 2 \pi r n^{*}(r)dr \nonumber \\
 &=\pi^{3/2}\frac{\xi\sigma_{0}}{m_{0}}\left(1\,au\right)^{2}\left(\frac{1}{2-n_{s}-\beta}\right)\times 
 \left(\left(\frac{r_{max}}{1\,au}\right)^{2-n_{s}-\beta}-\left(\frac{r_{min}}{1\,au}\right)^{2-n_{s}-\beta}\right)
\end{align}
where $r_{min}=2$ au, $r_{max}=10$ au. As in this study we are focusing on the role of OSS planetesimals, we will consider only the outer part of the protoplanetary disk, \textit{i.e.}, $r_{min}=4$ au and $n_{mp}=6\times10^{4}$.
%The same can be done for the results of simulations of Paper II using $n_{mp}=6\times10^{4}$, $r_{min}=2$ au and $r_{max}=8$ au.  %, \textit{i.e.}, $1\,au=1.49597870691\times10^{13}$ cm.\\
%
\subsubsection{Planetesimals Formed in a Turbulent Disk}
The second SFD considered is that of planetesimals formed by concentration of dust particles in low vorticity regions in a turbulent protoplanetary nebula \cite{cuzzi2008,cuzzi2010}. Following \cite{chambers2010}, the protoplanetary nebula is characterized by a surface density $\sigma'_{0}=3500$ g cm$^{-2}$ at $1$ au, a nebular density profile with exponent $n'_s=-1$ and a dust-to-gas ratio $\xi'=0.01$ beyond the Snow Line and $\xi'=0.005$ inside the Snow Line (see {Figure 14} from \cite{chambers2010}, gray dot-dashed line). Differently from Papers I and II and from \cite{chambers2010}, here the Snow Line will be placed at $4.0$ au. As in the case of the SFD by \cite{coradini1981}, the initial mass of solids contained in the region comprised between $2$ and $3$ au is about $4$ M$_{\oplus}$. The results of \cite{chambers2010} supply the average diameter of planetesimals as a function of heliocentric distance (see {Figure 14} from \cite{chambers2010}, gray dot-dashed line), from which Paper I derived the following semi-empirical relationship analogous to Equation  \eqref{masslaw}:
\begin{equation}\label{chambers_mass}
 \overline{m'}_{p}=\frac{\pi}{6}\rho D_{0}^{3}\left( \frac{r}{1\,au} \right)^{3\beta'}
\end{equation}
where $\beta'=0.4935$ and $D_{0}=70$ km is the average diameter of the planetesimals at $1$ au. By substituting the primed quantities to the original ones in Equations \eqref{massval} and \eqref{ntot}, the mass and the normalization factor for each massless particle can be obtained through the same approach described previously.\\
\subsubsection{The ``Asteroids were Born Big'' Scenario}
The third SFD considered is derived from the results of \cite{morbidelli2009}. Morbidelli \textit{et al.} \cite{morbidelli2009} did not explore a specific model of planetesimal formation in quiescent or turbulent disks but instead tried to constrain the initial \mbox{size-frequency} distribution of planetesimals in the orbital region of the asteroid belt, assuming an initial mass of $1.6$ M$_{\oplus}$ in the region comprised between $2$ and $3$ au. Their results suggest that the best match with the present-day SFD of the asteroid belt is obtained for planetesimal sizes initially spanning \mbox{$100$--$1000$ km} (see {Figure 8} from \cite{morbidelli2009}), a range consistent with their formation in a turbulent nebula. Accretion and break-up of these primordial planetesimals extended the size distribution between \linebreak $5$--$5000$ km (see {Figure 8a from \cite{morbidelli2009}}, black solid line) in about $3$ Ma, assumed by \cite{morbidelli2009} to be the formation time of Jupiter respect to CAIs.

The number of planetesimals in the orbital range between $4$ and $10$ au was estimated for this SFD by assuming a protoplanetary nebula similar to that described  for the SFD of \cite{coradini1981} but with a surface density $\sigma_{0}$ at $1$ au chosen so that the total mass of solids between $2$ and $3$ au is equal to $1.6$ M$_{\oplus}$. The population of planetesimals supplied by \cite{morbidelli2009} was then multiplied by the ratio between the mass contained in the 4--10 au orbital region to that of the 2--3 au orbital region. For each OSS impact event in the simulations of Paper I the mass of the impacting planetesimal was then estimated through a simple Monte Carlo extraction based on the cumulative probability distributions of the SFD supplied by \cite{morbidelli2009}. The normalization factor was estimated through Equation \eqref{ratio} using the scaled number of planetesimals.
\subsubsection{The ``Asteroids were Born Small'' Scenario}
The final SFD considered is derived from the results of \cite{weidenschilling2011}, who studied the accretion of primordial planetesimals in an annular region comprised between $1.5$ and $4$ au and containing $4.9$ M$_{\oplus}$ ($2$ M$_{\oplus}$ in the region between $2$ and $3$ au). Using an approach analogous to the one also used by \cite{morbidelli2009} but differing in the algorithms governing the computation of the collisional probabilities, \cite{weidenschilling2011} showed that a primordial SFD of the asteroid belt capable of reproducing the features of the present day SFD can be obtained also from disks initially populated by planetesimals as small as 50--200 m. Planetesimals of $500$ m in diameter succeed only partially in producing a satisfactory SFD. This study focuses on the SFD of the asteroid belt that \cite{weidenschilling2011} refers to as the ``standard case'', \textit{i.e.}, the one produced from a disk initially populated by planetesimals with a diameter of $100$ m (see {Figure 8} from \cite{weidenschilling2011}). 

As in the case of the SFDs by \cite{morbidelli2009}, the number of planetesimals in the orbital range between $4$ and $10$ au was estimated for this SFD by assuming a protoplanetary nebula similar to that described  for the SFD of \cite{coradini1981} but with a surface density $\sigma_{0}$ at $1$ au chosen so that the total mass of solids between \mbox{$2$ and $3$ au} is equal to $2$ M$_{\oplus}$. The mass of the impacting planetesimals was then estimated through a simple Monte Carlo extraction and the normalization factor $\gamma$ was computed using the scaled number of planetesimals. In the following, when using this SFD only planetesimals whose size is greater than or equal to $1$ km will be considered. %Note that in the size range considered, the SFD of the primordial asteroids would not change significantly should the initial mass of the region between $2$ au and $3$ au be raised to $4$ M$_{\oplus}$ (see Figure 19 in \cite{weidenschilling2011}).

%%%%%%%%%%%%%%%%%%%%%%%%%%%%%%%%%%%%%%%%%%

\subsection{Collisional Model: Cratering, Erosion and Water Delivery}\label{model-impacts}

%To estimate the fluxes of impactors on Vesta, Paper I opted for a statistical approach based on solving the ray--torus intersection problem between the orbital torus of Vesta and the linearised path of a massless particle across a time step. The method is similar to the analytical method developed by \cite{opik1976}, but does not require averaging over orbital angles other than the mean anomaly. Interested readers are referred to Papers I and II for details on the algorithm.

To estimate the fluxes of impactors on Vesta, Paper I opted for a statistical approach based on solving the ray--torus intersection problem between the orbital torus of Vesta and the linearized path of a massless particle across a time step. The method is similar to the analytical method developed by \cite{opik1976}, but does not require averaging over orbital angles other than the mean anomaly. Interested readers are referred to Papers I and II for details on the algorithm.

We used the fluxes estimated in Paper I as the basis for an improved assessment of the collisional evolution of Vesta during the JEB. A set of $10^{4}$ Monte Carlo simulations was run for each of the planetesimal SFDs described in Section \ref{model-sfds}. In each run a new mass value was extracted for each impact event recorded in the simulations of Paper I and, using the corresponding impact velocity, we computed the diameter and depth of the produced crater, the energy of the impact event, and the eroded mass. Averaging over each set of $10^{4}$ Monte Carlo simulations, we then computed for each SFD of the primordial planetesimals the total eroded mass, the cumulative probability of Vesta undergoing catastrophic disruption, and the fraction of the Vestan surface affected by the impacts.

The diameter of the craters produced by the flux of impactors was estimated, as in \cite{turrini2013}, using the following scaling law for rocky targets by \cite{holsapple2007}:
\begin{align}\label{crater_law}
\frac{R_{c}}{r_{i}}=0.93\left(\frac{g\,r_{i}}{v_{i}^{2}}\right)^{-0.22}	\left(\frac{\rho_{i}}{\rho_{v}}\right)^{0.31}
+0.93\left(\frac{Y_{v}}{\rho_{v} v_{i}^2}\right)^{-0.275}\left(\frac{\rho_{i}}{\rho_{v}}\right)^{0.4}
\end{align}
where $R_{c}$ is the final radius of the crater, $r_{i}$ is the radius of the impactor, $g=0.25$ m\,s$^{-2}$ is the surface gravity of Vesta, $v_{i}$ is the impact velocity, $Y_{v}=7.6$ MPa is the strength of the material composing the surface of Vesta (assumed to behave as soft rock, \cite{holsapple1993}), $\rho_{i}=1000$ kg/m$^3$ is the average density of the OSS impactors, and $\rho_{v}=3090$ kg/m$^3$ is the density of the basaltic surface of Vesta \cite{russell2012,russell2013}. 

%From the results of Equation \eqref{crater_law} we then estimated the excavation depths of the crater populations in the same way as \cite{turrini2013}. According to \cite{vincent2013}, the transition from simple to complex craters on Vesta seems to occur at diameters of about $30$ km. For craters smaller than this value we assumed a constant \mbox{depth-to-diameter} ratio of $0.168$, \textit{i.e.}, equal to the average value measured on Vesta by the Dawn \mbox{mission \cite{vincent2013}}. For larger craters we used the conservative relation from \cite{melosh1989}

From the results of Equation \eqref{crater_law} we then estimated the excavation depths of the crater populations in the same way as \cite{turrini2013}. According to \cite{vincent2013}, the transition from simple to complex craters on Vesta seems to occur at diameters of about $30$ km. For craters smaller than this value we assumed a constant \mbox{depth-to-diameter} ratio of $0.168$, \textit{i.e.}, equal to the average value measured on Vesta by the Dawn \mbox{mission \cite{vincent2013}}. For larger craters we used the conservative relation from \cite{melosh1989}
\begin{equation}\label{crater_depth}
d_{exc}=0.1\,D_{t}=0.1\left(D/1.3\right)=0.077\,D
\end{equation}
where $d_{exc}$ is the depth of excavation, $D_{t}$ is the diameter of the transient crater and $D$ is the final diameter of the crater. The factor $1.3$ is used to scale the diameter of the transient crater to that of the final \mbox{crater \cite{melosh1989,holsapple1993}.}

To assess how energetic the different impact events are and to which mass loss regime they are associated to, for each impact event we computed its specific kinetic energy per unit of the vestan mass $Q_{D}$ and compared it with the catastrophic disruption threshold $Q^{*}_{D}$ of Vesta. $Q^{*}_{D}$ is defined as the energy needed to destroy a target body, leaving a largest fragment with a mass equal to half the mass of the original body \cite{benz1999}. We evaluated the catastrophic disruption threshold $Q^{*}_{D}$ of Vesta using Equation ($6$) from \cite{benz1999} and the coefficients for basaltic targets computed by these authors (see Table $3$ from \cite{benz1999}). \linebreak As in Papers II and III, the coefficients of the case $v_{i}=5$ km\,s$^{-1}$ were used for all impact events with a velocity greater than or equal to $5$ km\,s$^{-1}$, and those of the $v_{i}=3$ km\,s$^{-1}$ were used for all the other impact events. For high-energy impacts ($0.1 \leq Q_{D}/Q^{*}_{D} < 1$), we computed the mass loss of Vesta using Equation ($8$) from \cite{benz1999} expressed in terms of the eroded mass $m_{e}$: 
\begin{equation}\label{erosion_benz}
\frac{m_{e}}{m_{t}}=0.5+s\left(\frac{Q_{D}}{Q^{*}_{D}}-1.0\right)
\end{equation}
where $m_{t}$ is the mass of the considered target body, $s=0.5$ for $v_{i}<5$~km\,s$^{-1}$ and $s=0.35$ for $v_{i} \geq 5$~km\,s$^{-1}$. The effects of catastrophic impacts ($Q_{D}/Q^{*}_{D} \geq 1$)  were not directly accounted for in the estimates of the eroded mass, as they would always result (by definition) in the loss of half the mass of Vesta. To properly treat them and include them in the collisional model, we would need to simulate the following dynamical evolution of the fragments and to estimate their re-accretion efficiency. The cumulative number of catastrophic impacts was used instead to assess the probability of Vesta surviving the JEB intact and to reject those scenarios where Vesta has a large chance of undergoing such events. Finally, the mass loss associated to cratering erosion caused by low-energy impacts ($Q_{D}/Q^{*}_{D} < 0.1$) was calculated based on the results of the hydrodynamic simulations we will now detail.

%%%%%%%%%%%%%%%%%%%%%%%%%%%%%%%%%%%%%%%%%%

In order to determine the amount of water and volatile elements delivered to Vesta by OSS impactors, we carried out a set of numerical simulations of the impacts of icy planetesimals on the asteroid. Numerical simulations of impacts on Vesta, treated as a planar target, have been previously performed by \cite{svetsov2011}, using a hydrodynamic approach and assuming that both the projectile and the target consisted of dunite with a density of $3.32$ g/cm$^3$. 

%With respect to \cite{svetsov2011}, in this study we made some modifications and improvements of the numerical model. As in \cite{svetsov2011}, we used the numerical hydrodynamic method SOVA (\cite{shuvalov1999}, SOVA is an acronym for Solid-Vapour-Air, as the code is designed for simulations of multi-material, multi-phase flows), but we adapted the hydrodynamic equations and the method to a 3D spherical system of coordinates.

With respect to \cite{svetsov2011}, in this study we made some modifications and improvements of the numerical model. As in \cite{svetsov2011}, we used the numerical hydrodynamic method SOVA (\cite{shuvalov1999}, SOVA is an acronym for Solid-Vapor-Air, as the code is designed for simulations of multi-material, multi-phase flows), but we adapted the hydrodynamic equations and the method to a 3D spherical system of coordinates.
This choice seems to be more convenient to study impacts that are not as large as the high-energy impacts that can shatter the target but at the same time are not so small that the sphericity of the target could be neglected. The gravitational field was assumed central and constant in time. A similar approach has been developed in 2D spherical geometry for simulations of large vertical impacts on the Earth \cite{svetsov2005}. Furthermore, we added the effects of dry friction as described in \cite{dienes1970}. The equations with friction are similar to the Navier-Stokes equations and dry friction does not violate the hydrodynamic similarity: the results, however, can depend on the dimensionless coefficient of friction. For the latter we adopted a value of $0.7$, which is typical for rocks and sand. The numerical grid consisted of $250\times100\times225$ cells over azimuth, polar angle and the radial distance respectively. We assumed bilateral symmetry, which allowed us to model only the half-space in the zenith direction (we assumed that the impact velocity vector lied in the reference plane that passes through the origin of coordinates and is orthogonal to the zenith). \linebreak Cell sizes were $1/40$ of the projectile's diameter around the impact point and increased to the antipodal point and to the radial boundaries located at distances of about $10$ vestan radii. The minimum cell sizes varied and became larger when the crater grew.

%\begin{table*}[t]
%\centering
%\begin{tabular}{cccc}
%ine
%Impact Velocity & Cometary Impactors & Asteroidal Impactors & Cometary Impactors \T\\ 
% (km/s)& (with friction) & (with friction) & (without friction) \B\\
%ine
%1	& 0.19 & 0.9	 &	0.23		\T\\
%2	& 0.49 & 2	&	0.82		\\
%4	& 3.8 & 4.65	&	2.84		\\
%6	& 6.5 &	8	&	6.3		\\
%8	& 11.5 & 	15	&	9.7 \\
%10	&	15	& 24.7	&	13.3\B\\
%ine
%\end{tabular}
%\caption{Erosion}\label{table-erosion}
%\end{table*}

Vesta was assumed to be in equilibrium and at a low temperature. Its radius was assumed equal to $260$ km, the core radius equal to $110$ km \cite{russell2012,russell2013}, the thickness of the crust equal to $23$ km, and the mass of Vesta equal to $2.59\times10^{23}$ g. We used the available ANEOS equations of state \cite{thompson1972} for the mantle and the crust with input data from \cite{pierazzo1997} and Tillotson's equation of state \cite{tillotson1962} for the iron core. ANEOS is an acronym for Analytic Equations of State: however, ANEOS does not have a simple analytic form but is instead a code with a large number of input parameters, the choice of which for a definite material is a difficult task. We assumed that the mantle material consists of dunite and the crust consists of granite. Usually just these materials are used in simulations of impacts because they are the only ones with proper equations of state that fit for the mantle and the crust. However, we should elucidate that various equations of state appropriately describe the state of materials compressed by the shock wave. The difficulties arise if one wants to obtain the amount of melted or vaporized material after the release phase. The impact velocities on Vesta are fairly low and the mass of escaped material, which is determined by the flow at rather low velocities, can be calculated with a reasonable accuracy for the chosen equations of state. We assumed that the icy planetesimals consisted of water and had a spherical shape. The diameter of impacting bodies in the simulations was set to 1 km: hydrodynamic similarity, however, holds valid when the diameters of the projectiles are much smaller than the diameter of Vesta and smaller than the thickness of the vestan crust. Therefore, the results of our simulations are valid for diameters of the impactors up to about $10$--$20$ km. The impact velocities ranged from $1$ to $10$ km/s and the impact angle was taken to be $45\degree$, which is the most probable impact angle \cite{melosh1989}.

%\begin{table*}[t]
%\centering
%\begin{tabular}{cccc}
%ine
%Impact Velocity & Cometary Impactors & Asteroidal Impactors & Cometary Impactors \T\\ 
% (km/s)& (with friction) & (with friction) & (without friction) \B\\
%ine
%1	&	0.45		&	0.91		&	0.54		\T\\
%2	&	0.33		&	0.66		&	0.51		\\
%4	&	0.25		&	0.45		&	0.42		\\
%6	&	0.09		&	0.38		&	0.28		\\
%8	&	0.03		&	0.32		&	0.13		\\
%10	&	0.03		&	0.29		&	0.09		\B\\
%ine
%\end{tabular}
%\caption{Retention}\label{fig-retention}
%\end{table*}

\begin{figure}[H]
\centering
\includegraphics[width=0.65\textwidth]{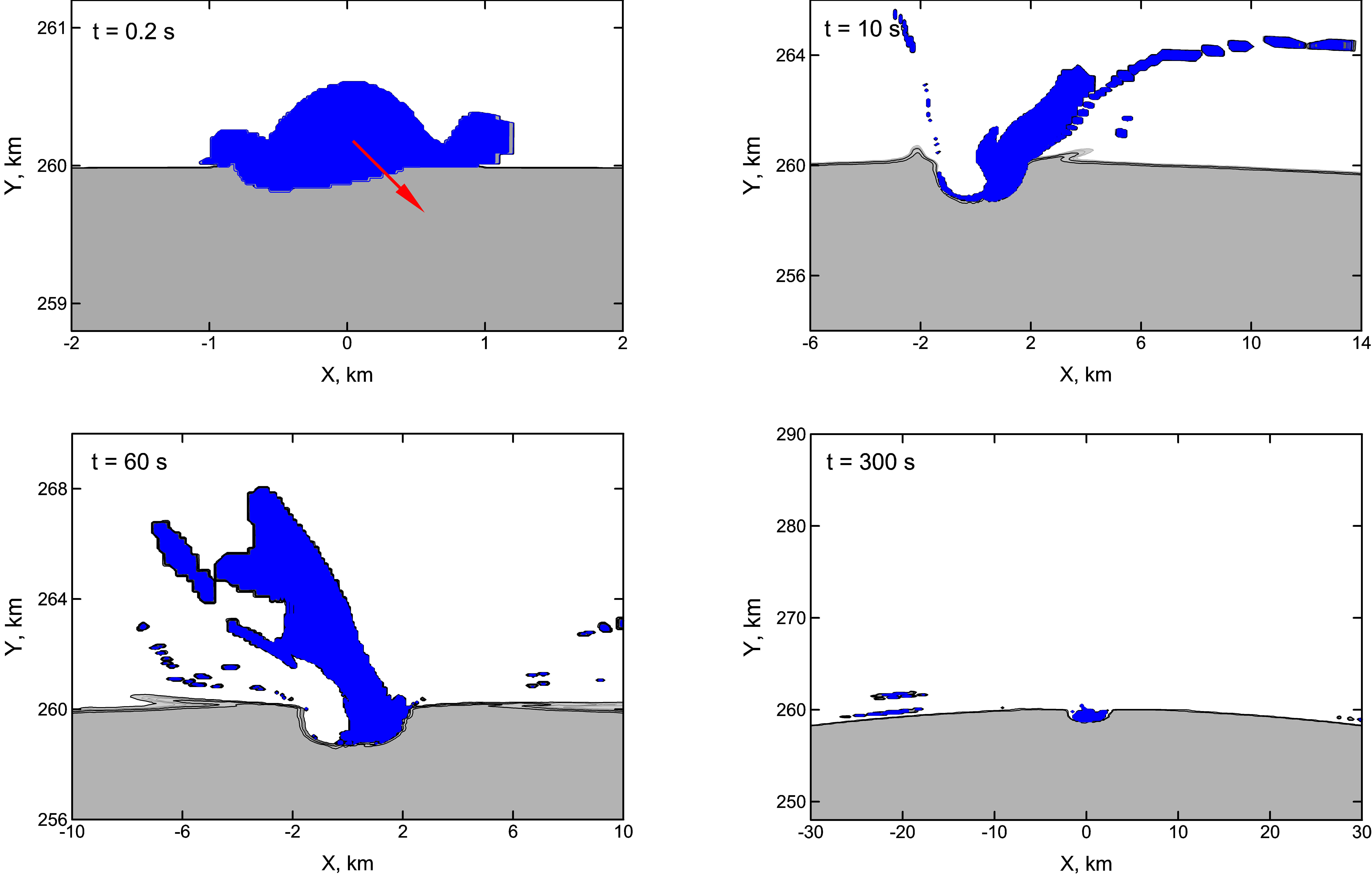}
\caption{Formation of the crater and evolution of the projectile's material after the impact on Vesta of a OSS planetesimal with a velocity 4 km/s and diameter of 1 km. Isolines of density are shown in the plane XY passing through the vector of impact velocity and the center of Vesta. The crustal material of Vesta with density above 0.5 g/cm$^3$ is shown by the grey color. The material of the impactor is indicated in blue. The red arrow in the first panel shows the direction of the impact (45$\degree$ respect to the local vertical). Note the different scales on the axes in the four panels.}\label{fig-crater}
\end{figure}

\begin{figure}[H]
\centering
\includegraphics[width=0.7\textwidth]{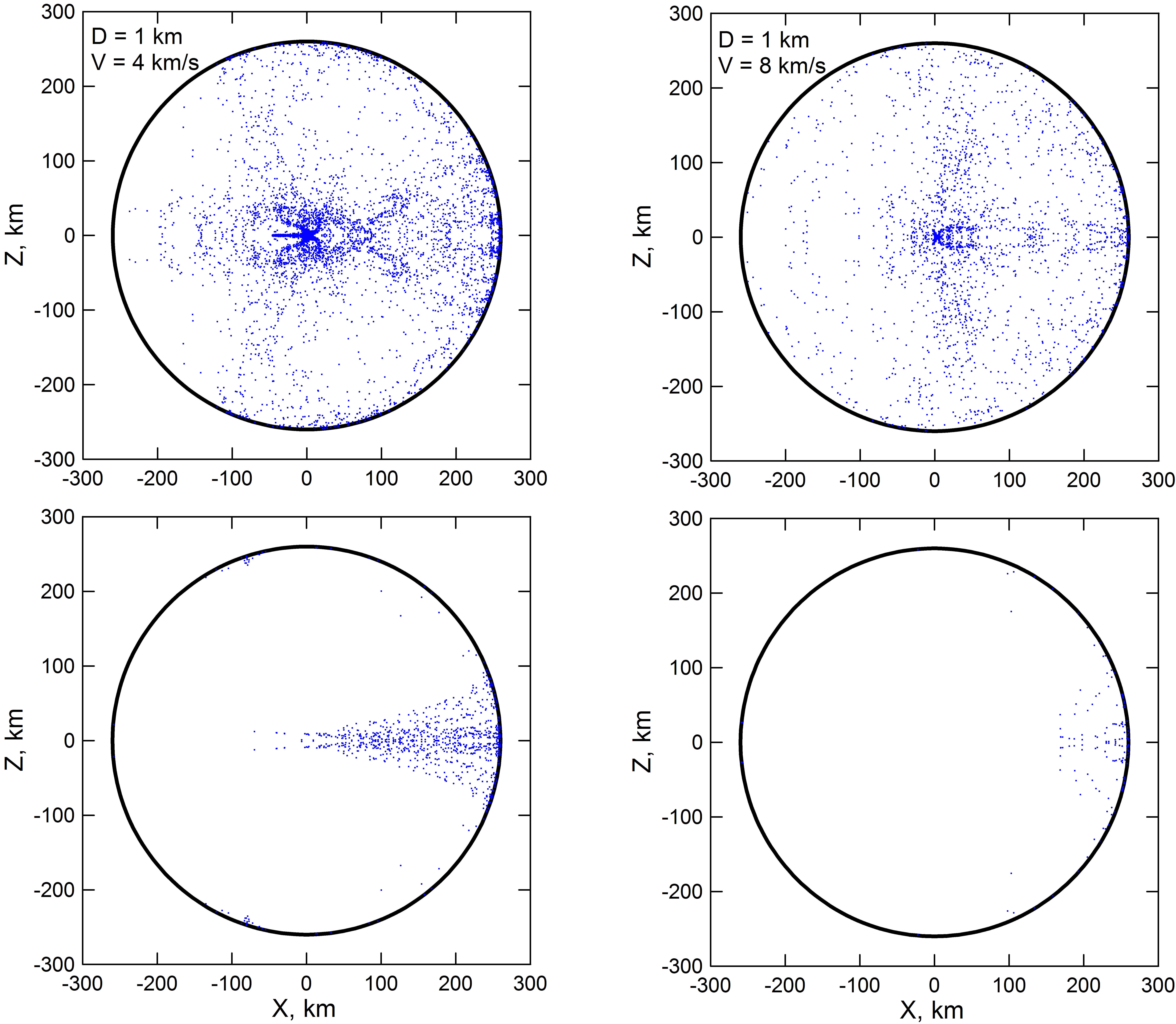}
\caption{Distributions of the material of the impactors over the vestan surface after the impacts of OSS planetesimals with diameter of 1 km  at 4 km/s (\textbf{left} plots) and at 8 km/s (\textbf{right} plots). The upper panels show the hemisphere where the impacts occurred, with the center of the craters in the middle of the hemisphere. The lower panels show instead the opposite hemisphere. The impacting bodies strike Vesta along the positive direction of the X axis at 45$\degree$ respect to the local vertical.}\label{fig-spreading}
\end{figure}

\begin{figure}[H]
\centering
\includegraphics[width=0.8\textwidth]{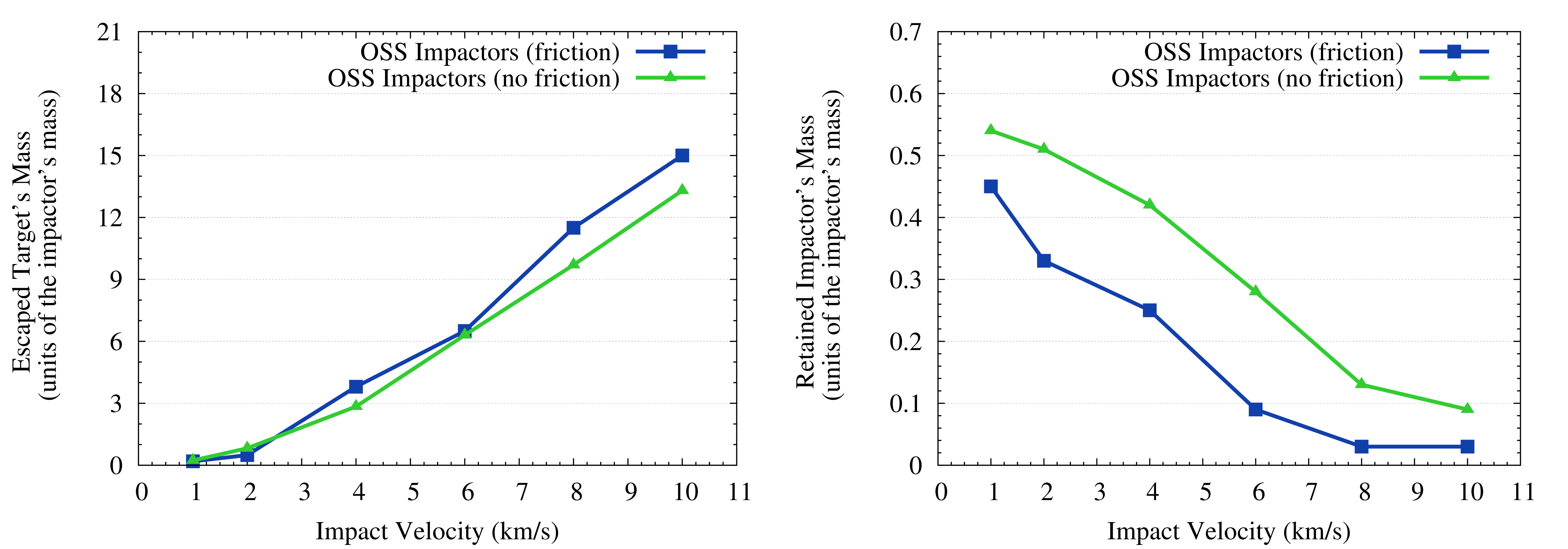}
\caption{Erosion (\textbf{left} plot) and accretion (\textbf{right} plot) efficiencies of cometary impactors on Vesta, expressed in units of the mass of the impactors. The plotted values are from Table \ref{table-sova}.}\label{fig-sova}
\end{figure}

During the impacts some portion of material ejected from the craters gained velocities higher than the escape velocity of Vesta ($0.35$ km/s). Obviously, the mass of the escaped material grows with the impact velocity. The retained material of the impacting planetesimals concentrates mainly within and around the impact craters, but small portions of the retained material are ejected and fall over the whole surface of Vesta as shown in Figures \ref{fig-crater} and \ref{fig-spreading}. The relative amounts of retained mass of the impactors and eroded mass of the target are reported in Table \ref{table-sova} and shown in {Figure \ref{fig-sova}}. The simulations of the impacts of icy bodies have been made both with and without friction. Friction consolidates the projectile's material and slightly accelerates the target's material. For probable impact velocities of $4$ km/s and \mbox{$8$ km/s} \mbox{(see Paper I)} the fraction of retained icy material is $0.25$ and $0.03$ of the projectile's mass. \linebreak If friction is neglected, these retained masses become $0.42$ and $0.13$ respectively. We used the values tabulated in Table \ref{table-sova} and shown in {Figure \ref{fig-sova}} also to estimate the erosion caused by the (low-energy) impacts of icy planetesimals. To do this, for each impact event recorded in the simulation of Paper I we derived the erosion efficiency associated to its impact velocity by linearly interpolating between the values tabulated in Table \ref{table-sova}.  

\begin{table}[H]
\centering
\begin{tabular}{ccccc}
\toprule

\textbf{Impact Velocity} & \textbf{Erosion} & \textbf{Erosion} & \textbf{Retention} & \textbf{Retention}\\ 
 \textbf{(km/s)} & \textbf{(with friction)} & \textbf{(without friction)} & \textbf{(with friction)} & \textbf{(without friction)} \\
\midrule
1	&	0.19	&	0.23	& 	0.45		&	0.54	\\
2	&	0.49	&	0.82	& 	0.33		&	0.51	\\
4	&	3.80	&	2.84	& 	0.25		&	0.42	\\
6	&	6.50	&	6.30	& 	0.09		&	0.28	\\
8	&	11.5	&	9.70	& 	0.03		&	0.13	\\
10	&	15.0	&	13.3	& 	0.03		&	0.09	\\
\bottomrule
\end{tabular}
\caption{Erosion and retention efficiencies of Vesta for cometary impacts, expressed in units of the mass of the impactors. The erosion and retention factors for the different impact velocities are computed both including and neglecting the effects of dry friction.}\label{table-sova}
\end{table}

Before proceeding, it must be noted that the ultimate fate of the retained water can depend on several processes and on their balance. First, water deposited in the crater can penetrate into the brecciated floor and be buried under the surface. Second, subsequent impacts can remove the water from the surface, vaporizing it (as suggested by \cite{denevi2012} as a possible origin of the pitted terrains on Vesta) or ejecting it with velocities higher than the vestan escape velocity. Third, material accreted by Vesta with low collision speeds as well as low-velocity ejecta from the craters of subsequent impacts can cover the retained water that remained on the surface and mix it with the target material. Forth, water lying or exposed on the surface will evaporate. Finally, in the temporal interval we are considering impacts can penetrate the solid crust of Vesta and deliver the water to the molten interior of the asteroid. Alternatively, impact-triggered effusive phenomena can cause the incorporation of part of the water on the surface into the solidifying magma.  In this study we will limit ourselves to the estimation of the maximum amount of water delivered to Vesta, neglecting the loss processes previously discussed.

%%%%%%%%%%%%%%%%%%%%%%%%%%%%%%%%%%%%%%%%%%

%To evaluate the surface erosion of Vesta, we proceeded in a way analogous to that of Paper II and Paper III. For low-energy impacts ($Q_{D}/Q^{*}_{D} < 0.1$), we used the scaling law for rocky targets by \cite{holsapple2007} in the form developed by \cite{svetsov2011} averaging over all impact angles:
%\begin{equation}\label{erosion_svetsov}
%\frac{m_{e}}{m_{i}}=0.03\left(\frac{v_{i}}{v_{V}}\right)^{1.65}\left(\frac{\rho_{i}}{\rho_{V}}\right)^{0.2}
%\end{equation}
%where $m_{e}$ is the escaped mass, $m_{i}$ is the mass of the impactor, $v_{i}$ is the impact velocity, $v_{V}$ is the escape velocity from the surface of Vesta and $\rho_{i}$ and $\rho_{V}$ are the densities respectively of the impactor and of Vesta.

\FloatBarrier

\section{Results}

To investigate the delivery of volatile elements to Vesta by the JEB, we will first characterize the flux of OSS impactors on the asteroid based on the results of Paper I. We will estimate the possibility of Vesta undergoing catastrophic impacts and we will quantify the erosion of its basaltic surface, to rule out implausible scenarios. We will then assess the amount of volatile elements that OSS impactors would deliver to the asteroid and, finally, the excavation of the surface due to the impacts, to assess whether the delivered volatile materials would be confined on the surface of the asteroid or could penetrate the solid crust and mix with the molten material. 

\subsection{Characterization of the Jovian Early Bombardment}\label{results-jeb}

\begin{figure}[H]
\centering
\includegraphics[width=0.85\linewidth]{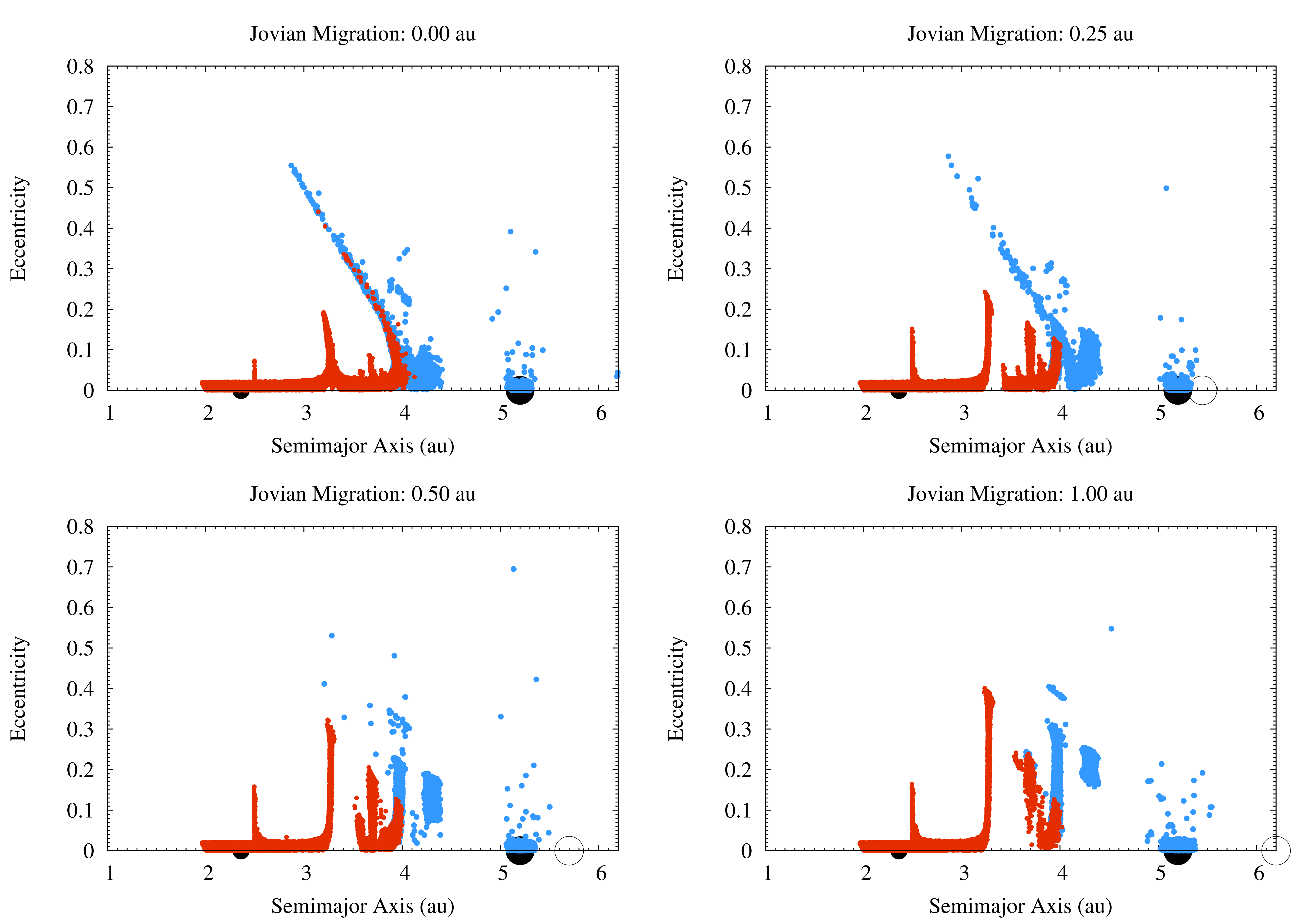}
\caption{The orbital structure of the protoplanetary disk $2\times10^{5}$ years after Jupiter started to accrete the nebular gas (\textit{i.e.}, $1.2\times10^{6}$ years from the beginning of the simulations). \protect\linebreak The black symbols at 2.36 au and at $5.2$ au indicate respectively Vesta and Jupiter. The open circles in the cases of 0.25 au, 0.50 au and 1.00 au migration indicate the initial position of Jupiter. ISS impactors are indicated in red, OSS impactors in blue.}\label{fig-jeb}
\end{figure}

\begin{figure}[H]
\centering
\includegraphics[width=0.75\linewidth]{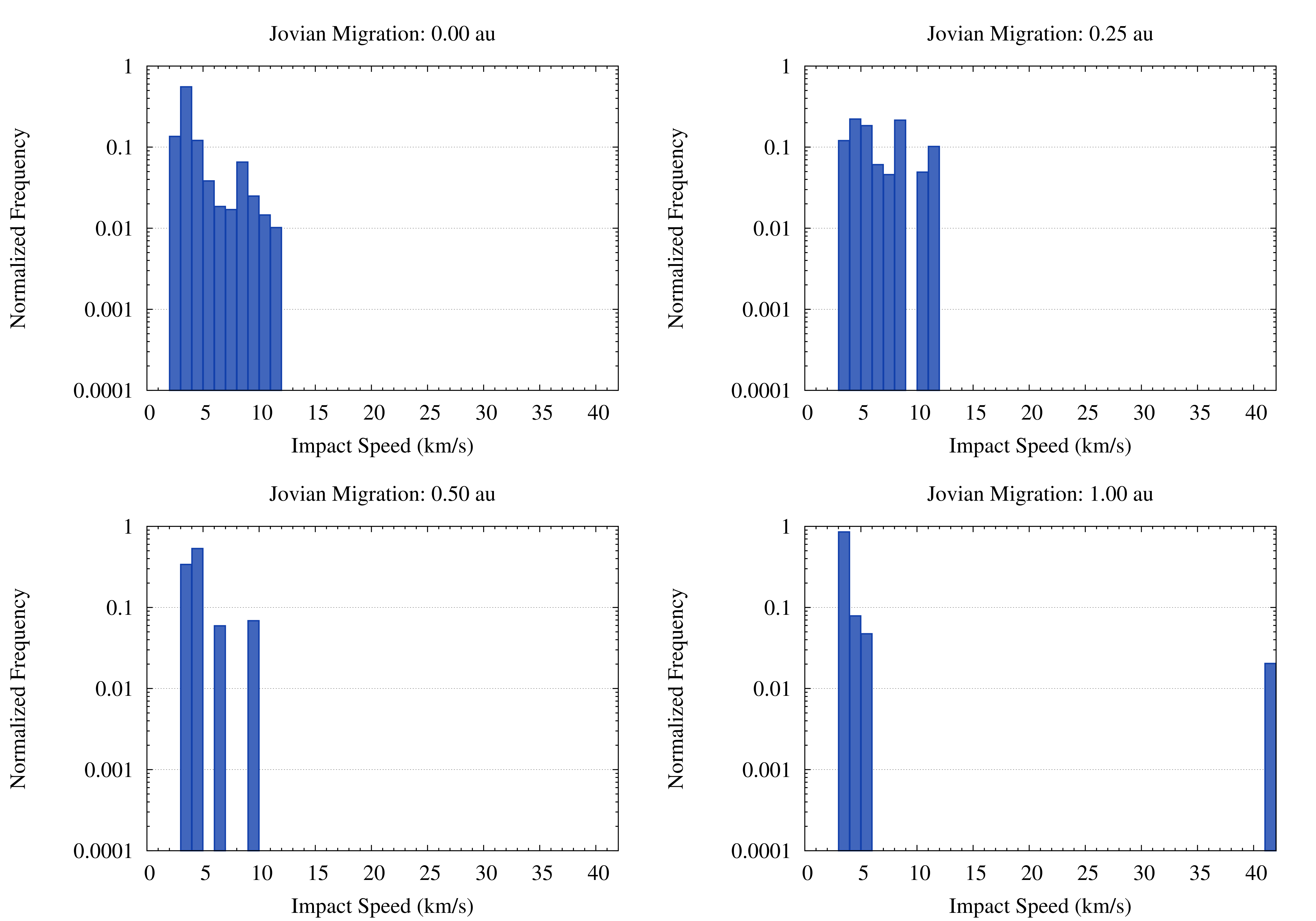}
\caption{Normalized distribution of the impact velocities of OSS impactors reported in Paper I in the four migration scenarios. The distribution of the impact velocities is common to all SFDs of the impactors.}\label{fig-speeds}
\end{figure}

The fluxes of OSS impactors hitting Vesta in the four migration scenarios we considered are quite different, as discussed in Paper I. The first difference is due to the distribution of the orbital elements of the OSS planetesimals in the semimajor axis--eccentricity (a--e) plane, as shown in {Figure \ref{fig-jeb}}, and divides the scenarios with limited (0.25 au) or no migration of Jupiter from the scenarios with a more significant displacement of the giant planet (0.5 au and 1 au). The different values of the orbital eccentricity of the impactors on Vesta between the migration scenarios (see {Figure \ref{fig-jeb}} and Figure 1 of Paper I) translate into a different distribution of their impact velocities, as shown in {Figure \ref{fig-speeds}}. In the scenarios with limited or no migration the impact velocities range from 2 km/s to 12 km/s, while in the scenarios with larger displacements of Jupiter the impact velocities mostly range between 3 km/s and 7 km/s. As reported already in Paper I, in the scenario where Jupiter migrates by 1 au about $2\%$ of the impacts occur at $\sim$$40$~km/s. %$$ 
For these impacts we considered that no fragment of the impactor survived on Vesta. 

\begin{figure}[H]
\centering
\includegraphics[width=0.75\linewidth]{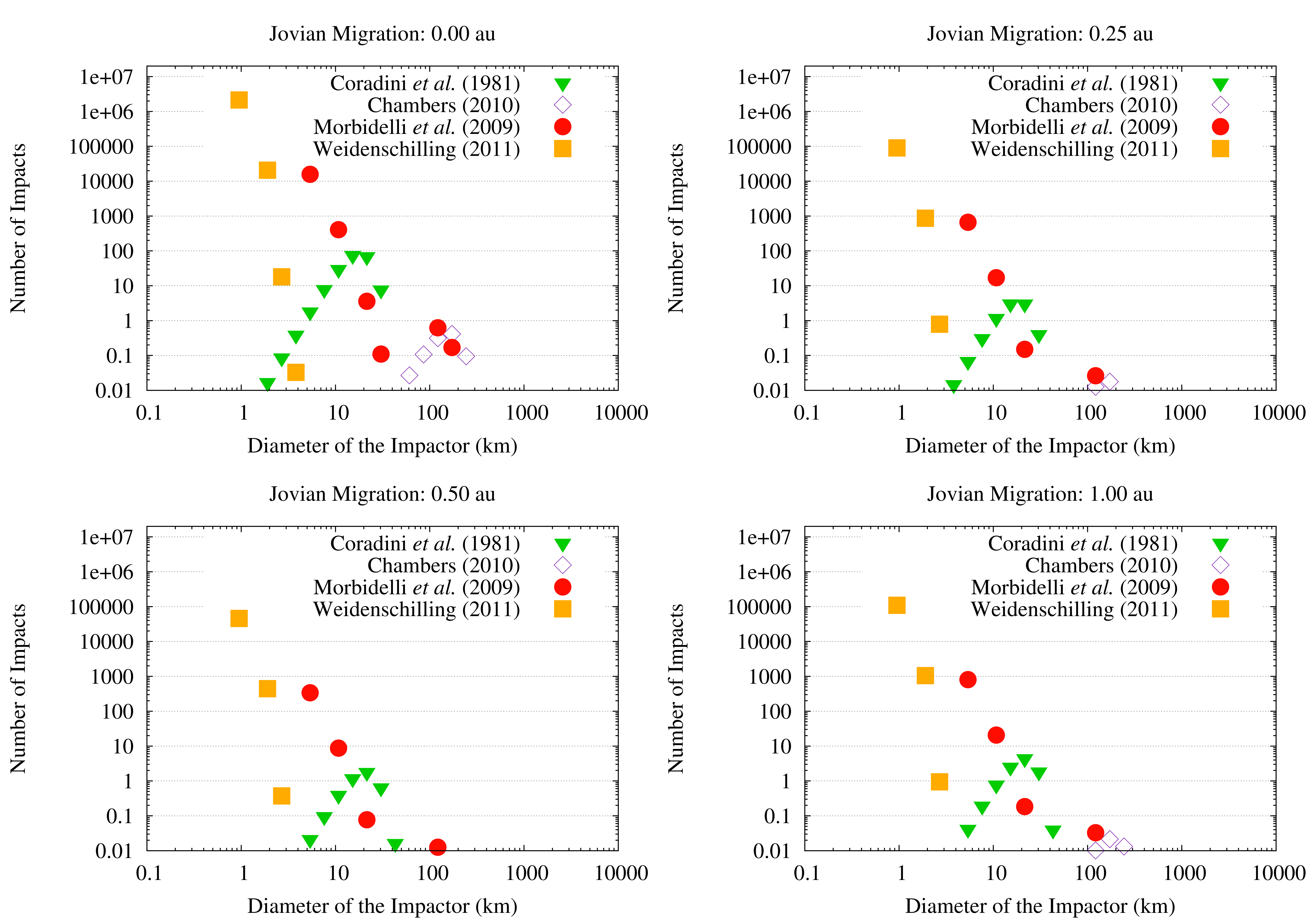}
\caption{Size-frequency distributions of the OSS impactors on Vesta in the four migration scenarios for the different SFDs of the primordial asteroids considered in this work. Numbers of impacts lower than 1 indicate stochastic events (even when considered cumulatively): these cases are not considered in the following analysis.}\label{fig-sizes}
\end{figure}

The second difference is in the number of OSS impactors on Vesta. As discussed in Paper I and here shown in {Figure \ref{fig-sizes}}, the scenario with no migration of Jupiter is the one with the highest flux of OSS impactors hitting the asteroid, the flux being 1--2 orders of magnitude higher than in the other cases. The number of OSS impactors on Vesta in the 0.25 au and 0.50 au migration scenarios decreases with the increase of the Jovian displacement, \textit{i.e.}, the larger the migration the lower the flux. The flux of OSS impactors then increases in the case of Jupiter migrating by 1 au: even if it still is 1 order of magnitude lower than in the case of no migration, the flux is the second highest among the scenarios we considered. Note that, because of this trend in the number of OSS impactors, the SFD by \cite{chambers2010} can produce (cumulatively) 1 impact only in the case where Jupiter does not migrate. 

As can be seen in {Figure \ref{fig-sizes}}, the OSS impactors are planetesimals whose diameter is about 1--2 km in the case of the SFD by \cite{weidenschilling2011}, about 5--10 km in the case of the SFD by \cite{morbidelli2009}, and about 10--30 km in the case of the SFD by \cite{coradini1981}. In the case of the SFD by \cite{chambers2010}, the OSS impactors that could hit Vesta would range in diameter between 100 km and 200 km but, as we discussed previously, this SFD can produce at most 1 impact on Vesta only in the case where Jupiter did not migrate.

\subsection{Catastrophic and High-Energy Impacts}\label{results-disruption}

As shown in Table \ref{table-disruption} and in agreement with the results of Paper I, the chances of Vesta being destroyed by a catastrophic impact during the JEB are always lower than $0.1\%$, mostly due to the limited duration of the JEB. The same holds true for high-energy impacts ($0.1\,Q_{D}^{*} \leq Q_{D} < Q_{D}^{*}$) in the cases of the SFDs by \cite{coradini1981,morbidelli2009,weidenschilling2011}, due to the relatively small size of the impactors (as discussed in Section \ref{results-jeb} and shown in {Figure \ref{fig-sizes}}). In the case of the SFD by \cite{chambers2010}, however, in the scenario where Jupiter does not migrate there is a fairly large chance ($\sim$$13\%$)  of Vesta undergoing a high-energy impact, which would have major consequences for the survival of its crust.
%$$

\begin{table}[H]
\centering
\begin{tabular}{ccccc}
\toprule

{\bf{Migration}} & {\bf{Coradini \textit{et al}.}} & {\bf{Chambers}} & {\bf{Morbidelli \textit{et al}.}} & {\bf{Weidenschilling}} \\ 

{\bf{(au)}} &{\bf{(1981)}} & {\bf{(2010)}} & {\bf (2009)} & {\bf{(2011)}} \\ 

\midrule

%0.00	&	0	&	0.00012831	&	0.00077372	&	1.3364e-10       \\
%0.25	&	0	&	5.5416e-05	&	8.3249e-05	&	2.648e-08        \\
%0.50	&	0	&	1.8521e-06	&	2.4044e-05	&	0                \\
%1.00	&	0	&	0.00092705	&	0.00051209	&	0                \\
0.00	&	0	&	$<$$10^{-3}$	&	$<$$10^{-3}$	&	$<$$10^{-3}$       \\
0.25	&	0	&	$<$$10^{-3}$	&	$<$$10^{-3}$	&	$<$$10^{-3}$       \\
0.50	&	0	&	$<$$10^{-3}$	&	$<$$10^{-3}$	&	0                \\
1.00	&	0	&	$<$$10^{-3}$	&	$<$$10^{-3}$	&	0                \\
\bottomrule

\end{tabular}
\caption{Number of catastrophic impacts on Vesta across the JEB in the different migration scenarios and for the different SFDs considered.}\label{table-disruption}
\end{table}

\subsection{Erosion of Vesta due to OSS Impactors}

\begin{figure}[H]
\centering
\includegraphics[width=0.6\linewidth]{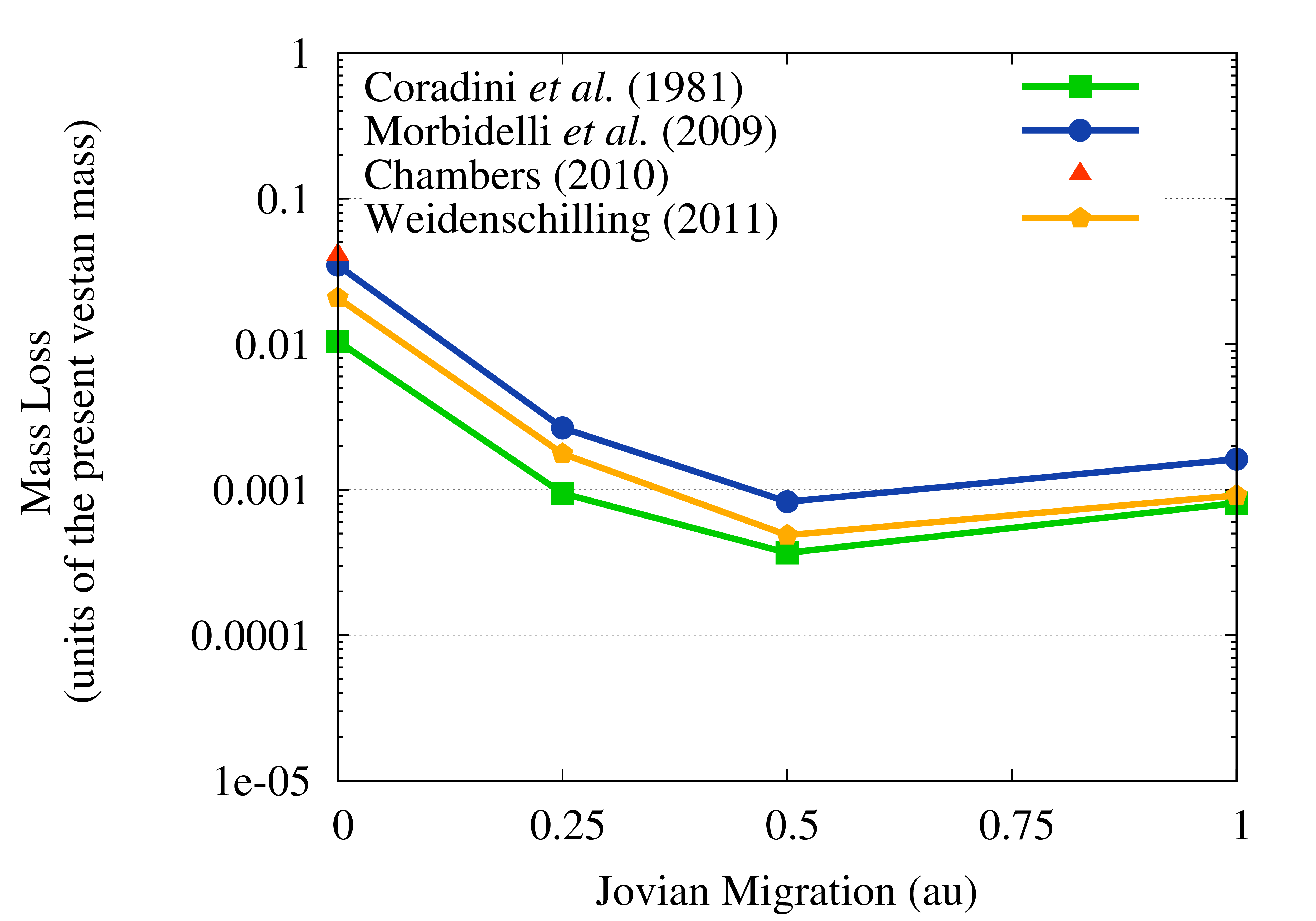}
\caption{Mass loss of Vesta in the different migration scenarios and for the different SFDs considered. The mass loss is expressed in units of the present Vestan mass (\cite{russell2012,russell2013}).}\label{fig-massloss}
\end{figure}

Now that we have verified that Vesta would survive the flux of OSS impactors triggered by the JEB without being shattered or destroyed, we need to assess the degree of erosion that these impactors would cause to its surface and test whether any of our scenarios is inconsistent with the survival of the basaltic crust of Vesta. In {Figure \ref{fig-massloss}} we show the mass loss of Vesta due to cratering erosion: as we expected following our discussion in Section \ref{results-disruption}, the case of SFD by \cite{chambers2010} and no migration of Jupiter results in a significant mass loss of the asteroid ($\sim$$4\%$), about twice as large as the one caused by ISS impactors as estimated in Paper III. 

%$$

Also in the case of the SFDs by \cite{coradini1981} the erosion caused by OSS impactors in the no migration scenario is twice as large as that caused by ISS impactors and estimated in Paper III, reaching about $1\%$ of the present vestan mass. In the cases of the SFDs by \cite{morbidelli2009,weidenschilling2011}, the erosion caused by the OSS impactors is respectively about six and twenty times larger than the one caused by ISS impactors in Paper III. The eroded mass amounts to about $3\%$ of the present vestan mass for the SFD by \cite{morbidelli2009} and to about $2\%$ of the present vestan mass for the SFD by \cite{weidenschilling2011}. In all the other migration scenarios, the cratering erosion caused by OSS impactors is significantly lower than the one due to ISS impactors and never rises above $0.3\%$ of the present vestan mass. As a consequence, the results of Paper III are not affected by the OSS planetesimals in those scenarios where Jupiter migrates.

To better understand the effects of the OSS impactors during the JEB for the survival of the basaltic crust of Vesta, we converted the mass loss values shown in Figure \ref{fig-massloss} to the thickness $\Delta R$ of the shell, extending outward from the present surface of Vesta, whose mass matches the mass $\Delta M$ lost by the asteroid, \textit{i.e.},
\begin{equation}
\Delta R = \left( R_{V}^{3} + \frac{3\Delta M}{4\pi\rho_{V}} \right)^{1/3} - R_{V}
\end{equation}
where $R_{V}=262.7$ km and $\rho_{V}=3090$ kg m$^{-3}$ are respectively the mean radius and the mean crustal density of Vesta measured by the Dawn mission (\cite{russell2012,russell2013}). %Before discussing the surface erosion of Vesta, it should be stressed that the values here computed for the thickness of the eroded layer cannot be compared directly with the thickness of the eucritic and diogenitic layers estimated by \citet{ruzicka1997}. As mentioned previously, in fact,  \citet{ruzicka1997} assumed that Vesta originally had a mean radius of $265$ km, so a direct comparison is possible only in cases where the surface erosion is of about $2-3$ km.

\begin{figure}[H]
\centering
\includegraphics[width=0.8\linewidth]{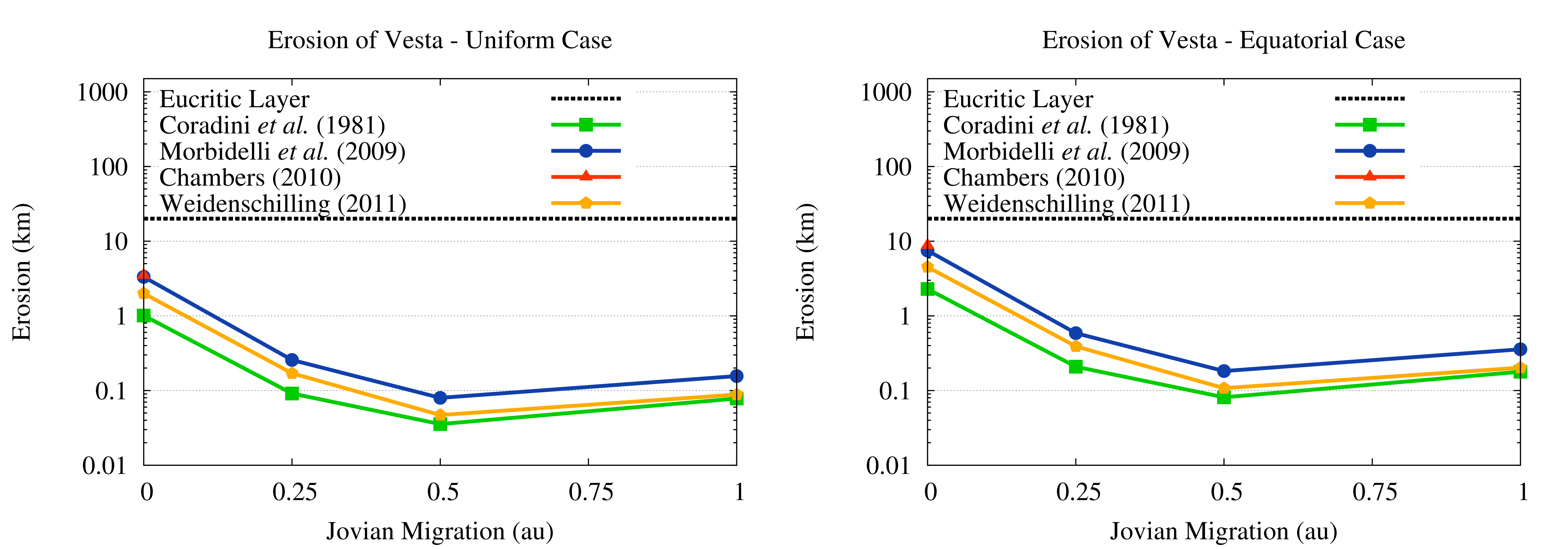}
\caption{Surface erosion of Vesta in the different migration scenarios and for the different SFDs considered. Surface erosion is expressed as the thickness of a shell, with density equal to the present crustal one of Vesta and mass equal to the eroded mass, extending from the present radius of Vesta outwards. The plot on the left shows the case of a uniform erosion, the plot on the right shows instead the case of erosion proportional to the cross-sectional area of Vesta. The dotted line indicates the thickness of the eucritic layer according to \cite{ruzicka1997}.}\label{fig-erosion}
\end{figure}

The values of the thickness of the eroded crust we obtained are shown in Figure \ref{fig-erosion}, where we considered both cases of erosion discussed in Paper III. The first one is the case of uniform erosion, which assumes that impacts are distributed isotropically on the surface of Vesta. The second case instead assumes, based on the compact vertical extension of the protoplanetary disk discussed in Section \ref{model-nebula}, that impacts on Vesta takes place mostly on the ecliptic plane and therefore they are distributed as a function of the cross-sectional area of the different regions of the asteroid. As a consequence, the region comprised between $45\degree$ and $-45\degree$ of latitude should receive about $70\%$ of the impacts and erode more than the regions between $-45\degree$ and $-90\degree$ and between $45\degree$ and $90\degree$. As discussed in Paper III, this case results in an erosion of this ``equatorial'' belt twice as high as the one we obtain in the isotropic case.

As we can see from Figure \ref{fig-erosion}, the erosion of the vestan crust in the isotropic case is of about \mbox{100--300 m} or less for all the SFDs in all the scenarios where Jupiter migrates while accreting the gas. In the scenario of no migration of Jupiter, the surface erosion of Vesta rises instead to 1--2 km \mbox{(see Figure \ref{fig-erosion})} for the SFDs from \cite{coradini1981,weidenschilling2011} and to 3 km for the SFDs from \cite{chambers2010,morbidelli2009}. Once we sum the erosion due to OSS impactors to the one we estimated for ISS impactors in Paper III, we obtain the erosion pattern discussed in Papers I and II. The erosion of Vesta decreases when we move from the scenario of no migration to that of limited (0.25 au) migration due to the drop in the flux of OSS impactors (characterized by higher velocities), which is not compensated by the increase in the ISS impactors (characterized by lower velocities). However, if Jupiter undergoes more extensive migration (0.50--1.00 au) the new ISS impactors due to the 2:1 resonance cause the erosion of Vesta to grow again, reaching values higher than those due to the ISS and OSS impactors in the no migration scenario.

Finally, in order to assess whether erosion would be a global effect on Vesta or not, we built the R-plot of the crater populations produced by the OSS impactors for the different SFDs of the planetesimals and the four migration scenarios. Following  \cite{melosh1989}, we took as the smallest crater diameter the value of $1$ km and we divided the crater population associated to each SFD in bins where $D_{i+1}=\sqrt{2}D_{i}$ and the central diameter is the geometric mean $\overline{D_{i}}=\sqrt{D_{i}D_{i+1}}$. We computed the R-value as
\begin{equation}
R_{i}=3.65 f_{i}
\end{equation} 
where $f_{i}$ is the fraction of the vestan surface covered by the craters in the relevant bin. The value of $f_{i}$ of each bin is obtained simply by summing the surface areas $A_{j}$ covered by the $N_{i}$ craters in the bin and dividing it by the surface of Vesta $S_{V}$:
\begin{equation}
f_{i}=\left( \Sigma_{j=0}^{N_{i}} A_{j} \right)/S_{V}
\end{equation} 
where $A_{j}$ is the geometrical area of each crater of diameter $D_{j}$:
\begin{equation}
A_{j}=\frac{\pi}{4}D_{j}^2
\end{equation}
Note that, when the condition $D_{i+1}=\sqrt{2}D_{i}$ is satisfied and the average diameter of each bin is computed as the geometric mean $\overline{D_{i}}=\sqrt{D_{i}D_{i+1}}$, this definition of the R-value is equivalent to the one given \mbox{by \cite{crater1978}}, \textit{i.e.},
\begin{equation}
R_{i}=\frac{N_{i}\,\overline{D_{i}}^{3}}{S\left(D_{i+1}-D_{i}\right)}
\end{equation}
where $N_{i}$ is the number of craters in the relevant bin and $S$ is the surface area whose population of craters is under study, which in our case is the whole surface of Vesta ($S=S_{V}$).
In building the R-value distributions, we considered only the effects of low-energy impacts ($Q < 0.1\,Q^{*}_{D}$): high-energy and catastrophic impacts were not included. Moreover, we considered only those bins where the cumulative impact probability produced at least $1$ impact once normalized to the real population of the disk. As a consequence, we did not consider the SFD by \cite{chambers2010}.

\begin{figure}[H]
\centering
\includegraphics[width=0.8\linewidth]{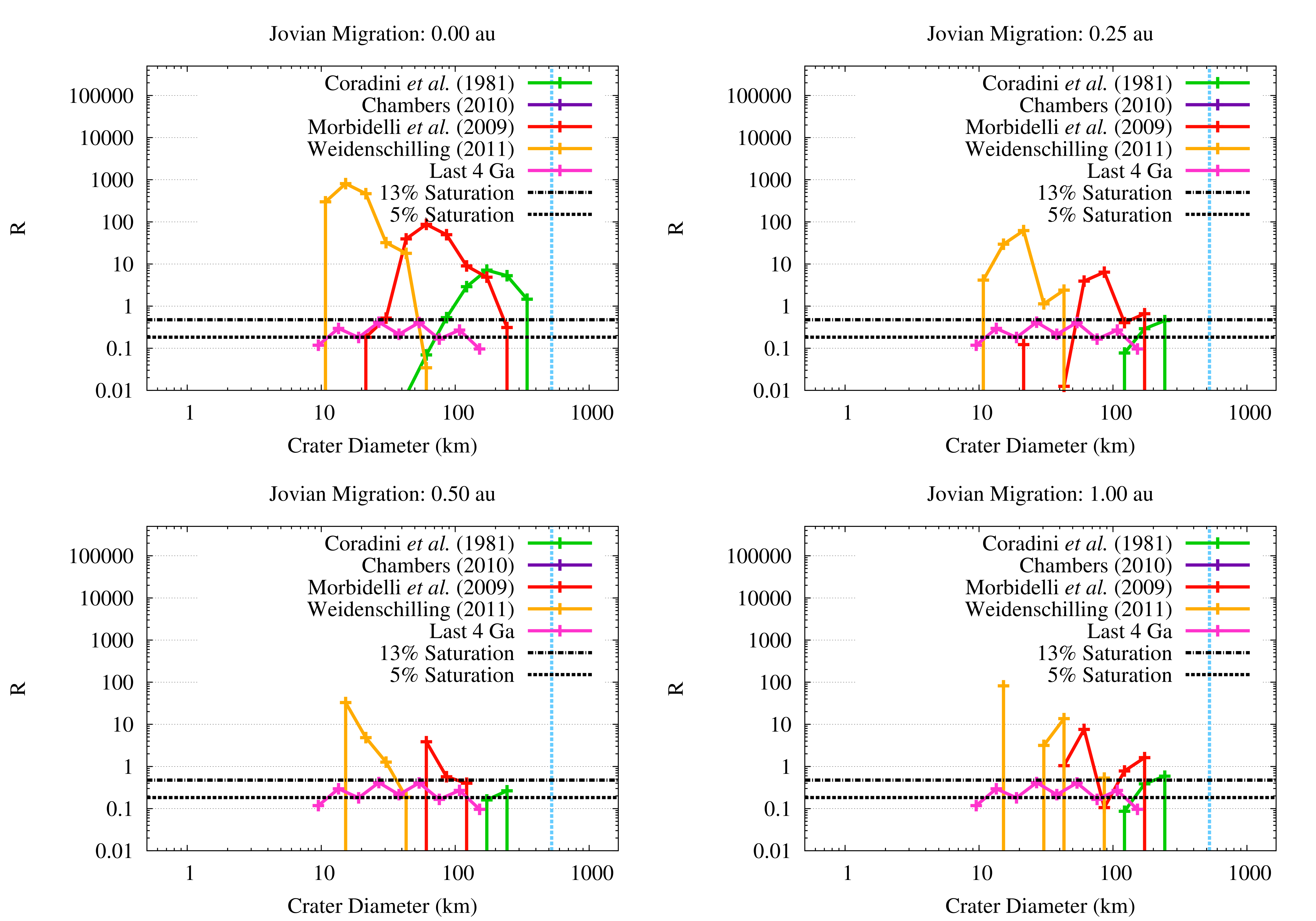}
\caption{R-plots of the crater populations produced by the different SFDs of the OSS planetesimals in the four migration scenarios. The $5\%$ and $13\%$ saturation levels and the crater population caused by the last $4$ Ga of collisional evolution of Vesta are also shown for reference. The light blue vertical dashed line indicates the present diameter of Vesta.}\label{fig-rplot}
\end{figure}

The R-plots for the SFDs by \cite{coradini1981,morbidelli2009,weidenschilling2011} are shown in Figure \ref{fig-rplot}, where we also showed the {R}-value that would be associated to a $5\%$ and a $13\%$ saturation of the surface of the asteroid. These two threshold levels represent respectively the minimum R-value for which a crater population can reach \mbox{equilibrium \cite{melosh1989}} and the {R}-value estimated for Mimas, whose surface is the most densely cratered in the Solar \mbox{System \cite{melosh1989}}. As a reference, in Figure \ref{fig-rplot} we also showed the {R}-values of the crater population produced on Vesta by asteroidal impactors over the last 4 Ga estimated by \cite{turrini20XX}. 

The crater population produced by the SFD from \cite{weidenschilling2011} mainly ranges in diameter between 10 km and 40 km. The impacts occurring at about 40 km/s results in craters with diameters of 90 km. The bulk of the crater population associated to the SFD from \cite{morbidelli2009} ranges in diameter between 60 km and 200 km, with a tail down to 20 km in diameter in the no migration scenario. Finally, the crater population due to the SFD by \cite{coradini1981} ranges between 40 km and 300 km in the scenario where Jupiter does not migrate, while in the scenarios where migration occurs the crater population concentrates between 100 km and 200 km.
As can be easily seen, even if the flux of OSS impactors is more limited with respect to the one of the ISS impactors, it is still large enough to saturate the surface of Vesta at least to a $10\%$ level. %comprised between $5\%$ and $13\%$. 
\mbox{The cratering} and the erosion produced by the OSS impactors therefore appear to globally affect the surface of Vesta.

\subsection{Delivery of Volatile Materials to Vesta}\label{results-water}

Now that we concluded the assessment of the implications of the flux of OSS impactors for the survival of Vesta and of its basaltic crust, we can move to investigating the delivery of water and volatile materials to the asteroid by the JEB. Before proceeding it is important to note that, while water realistically represented the majority of the delivered material, a varying fraction of the mass of the impacting OSS planetesimals was in the form of other ices (mainly carbon monoxide and carbon dioxide, see \cite{mumma2011}) and trace compounds (e.g., ammonia and methane, see \cite{mumma2011}). However, the available observational constraints are based on the detected emissions of present day comets \cite{mumma2011}, which are not necessarily representative of the primordial planetesimals considered in this study or even of the bulk composition of the very same comets. Moreover, the abundances of the different volatile materials vary significantly from one comet to the other \cite{mumma2011}, introducing another large source of uncertainty. Based on the observational evidences, water plausibly represented at least $50\%$ of the material delivered by OSS impactors, so we can constrain the uncertainty on the water delivery to about a factor of two.

The absolute values of the retained masses of volatile materials that we estimated with our model in the four migration scenarios and for the four SFDs of the planetesimals are reported in Table \ref{table-water}. As we can see, the amounts of water delivered to Vesta vary between a few $10^{18}$ g and a few $10^{20}$ g and, for a given migration scenario, the retained amount of water varies by a factor of $2$--$3$ between the lowest and the highest value. As a reference, the content of hydrogen and hydrated materials measured by the Gamma Ray and Neutron Detector (GRaND) on-board the Dawn spacecraft in the topmost 1 m of the vestan regolith \cite{prettyman2012} and suggested to be due to the cumulative flux of impactors on Vesta over the last \linebreak 4 Ga \cite{turrini20XX} is about $2.7\times10^{15}$ g. In the case of the SFD by \cite{chambers2010} the single value reported refers to the case of no migration of Jupiter, as in the other cases the cumulative impact probability of all OSS impactors is not enough to produce even a single impact. It must be noted, however, that the single impact associated to this SFD would produce major effects on the surface of Vesta and would likely destroy the basaltic crust, if it was already formed.

\begin{figure}[H]
\centering
\includegraphics[width=0.5\linewidth]{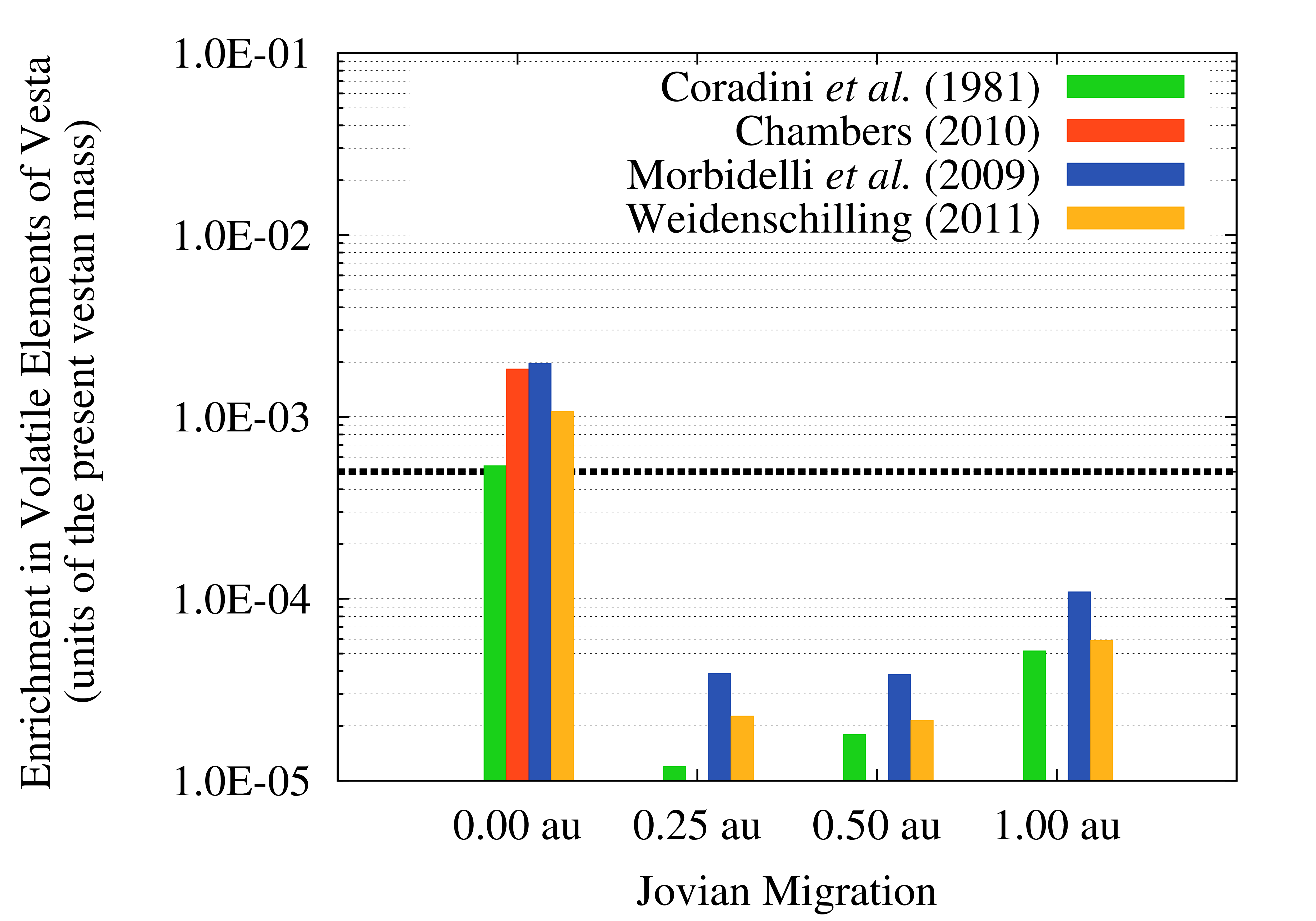}
\caption{{Amounts of volatile materials} (expressed in units of the present vestan \mbox{mass, \cite{russell2012})} delivered by the OSS impactors to Vesta in the four migration scenarios and for the four SFDs we considered. Also indicated, as the horizontal dotted line, is the fraction of the Earth's mass represented by water ($5\times10^{-4}$, \cite{morbidelli2000}).}\label{fig-water}
\end{figure}

In order to better illustrate the effects of the JEB in terms of the delivery of water and volatile materials to Vesta, in Figure \ref{fig-water} we show the quantities of Table \ref{table-water} expressed in units of the present mass of Vesta. If we compare these enrichment factors with the fraction of the Earth's mass represented by water, \textit{i.e.}, $5\times10^{-4}$ \cite{morbidelli2000}, we can immediately see that the scenario of no migration of Jupiter could result in a primordial Vesta as enriched in water as the Earth. In all the other migration scenarios, Vesta would be about one order of magnitude less enriched. As we mentioned in Section \ref{model-impacts} not all this volatile material delivered to Vesta will survive to the subsequent collisional evolution of the asteroid. In particular, as discussed in Paper III the flux of ISS impactors caused by the JEB is intense enough to strip Vesta of the equivalent of at least a few hundreds meters (if not a few kilometres) of surface material. It is therefore plausible that the fraction of volatile elements delivered during the JEB that remained nearer to the surface of the asteroid was removed by the JEB itself through the more abundant ISS impactors. However, if the impacts of the OSS planetesimals can excavate the solid crust of Vesta, part of the volatile material they deliver could be trapped into the molten interior and be incorporated into the crystallizing minerals as in the samples studied by \cite{sarafian2013}. As discussed by \cite{formisano2013,tkalcec2013}, in fact, depending on the accretion time and the initial porosity of Vesta the thickness of its solid crust across the JEB could have ranged from a minimum of $7$--$10$ km to a maximum of $20$--$30$ km.

\begin{table}[H]
\centering
\begin{tabular}{ccccc}
\toprule

{\bf Migration} & {\bf Coradini \textit{et al}.} & {\bf Chambers} & {\bf Morbidelli \textit{et al}. } & {\bf Weidenschilling} \\

{\bf (au)} & {\bf (1981)} & {\bf (2010)} & {\bf (2009)} & {\bf (2011)} \\

\midrule

0.00	&	$1.4\times10^{20}$ g	&	$4.7\times10^{20}$ g	&	$5.1\times10^{20}$ g	&	      $2.8\times10^{20}$ g \\
0.25	&	$3.0\times10^{18}$ g	&	-	&	$1.1\times10^{19}$ g	&	      $5.9\times10^{18}$ g   \\
0.50	&	$4.7\times10^{18}$ g	&	-	&	$1.0\times10^{19}$ g	&	      $5.5\times10^{18}$ g   \\
1.00	&	$1.3\times10^{19}$ g	&	-	&	$2.8\times10^{19}$ g	&	      $1.5\times10^{19}$ g \\
\bottomrule

\end{tabular}
\caption{Amounts of volatile materials delivered by OSS impactors to Vesta in the four migration scenarios and for the four SFDs we considered. As a reference value, to produce the same abundance of water of the Earth \cite{morbidelli2000}, OSS impactors should deliver $1.3\times10^{20}$ g.}\label{table-water}
\end{table}

\begin{figure}[H]
\centering
\includegraphics[width=0.75\linewidth]{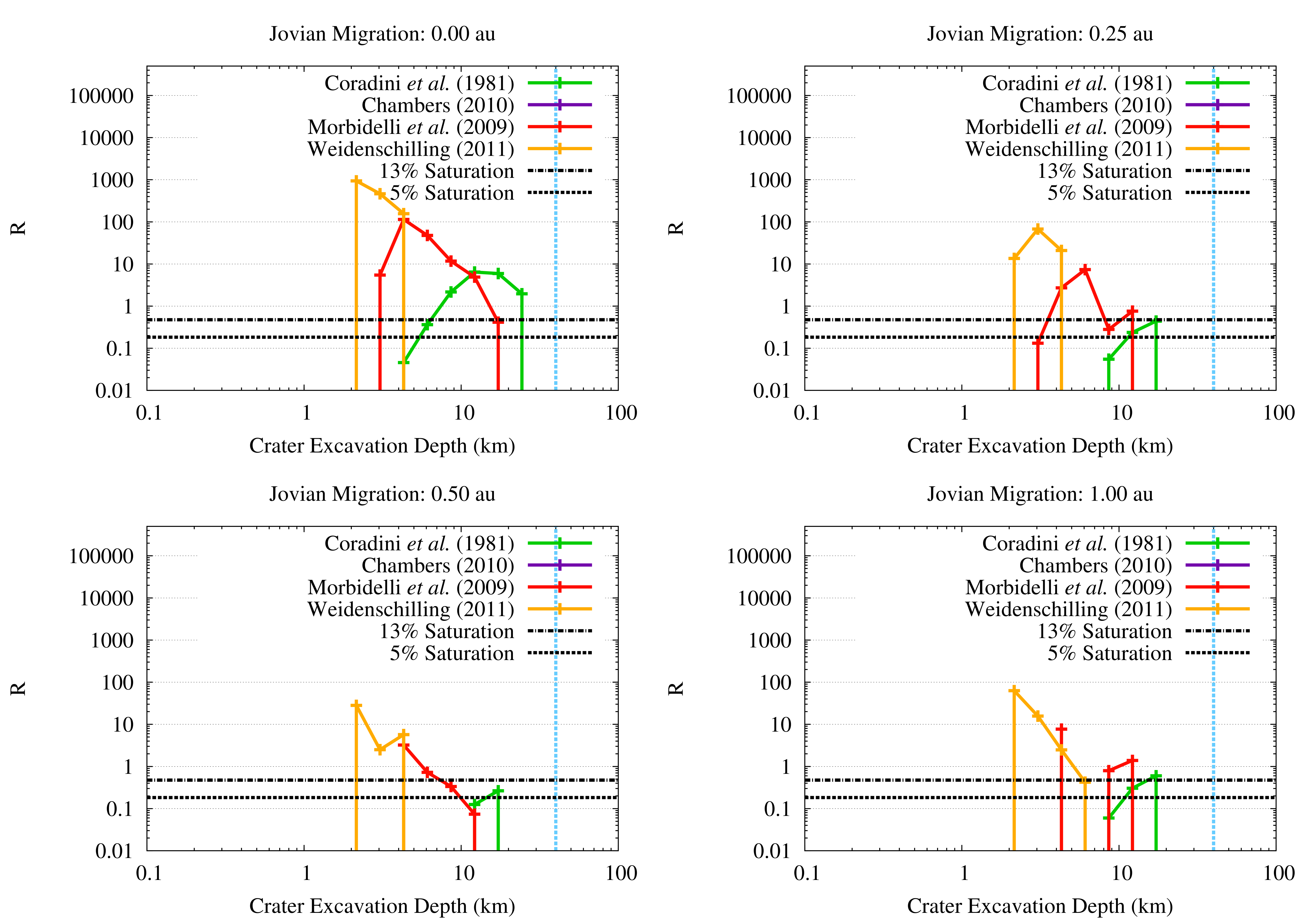}
\caption{R-plot of the excavation depth of craters on Vesta across the JEB for the different SFDs and migration scenarios considered in this work. The $5\%$ and $13\%$ saturation levels are shown for reference. The vertical light blue dashed line indicates the thickness of the crust of Vesta (\textit{i.e.}, the eucritic and diogenitic layers) as estimated by \cite{ruzicka1997}.}\label{fig-dplot}
\end{figure}

In order to assess whether the OSS impactors on Vesta would be capable of penetrating the solid crust and reach the molten interior, we estimated the excavation depths of the crater populations we discussed previously. %According to \cite{vincent2013}, the transition from simple to complex craters on Vesta seems to occur at diameters of about $30$ km. For craters smaller than this value we assumed a constant depth-to-diameter ratio of $0.168$, \textit{i.e.}, equal to the average value measured on Vesta by Dawn (\cite{vincent2013}). For larger craters we used the conservative relation from \cite{melosh1989}
%\begin{equation}\label{crater_depth}
%d_{exc}=0.1\,D_{t}=0.1\left(D/1.3\right)=0.077\,D
%\end{equation}
%where $d_{exc}$ is the depth of excavation, $D_{t}$ is the diameter of the transient crater and $D$ is the final diameter of the crater. The factor $1.3$ is used to scale the diameter of the transient crater to that of the final crater (\cite{holsapple1993}). 
Using the results of Figure \ref{fig-rplot}, we expressed the excavation of the vestan surface in the form of an R-plot of the crater depths across the JEB, as shown in Figure \ref{fig-dplot}. As we can see from \mbox{Figure \ref{fig-dplot},} the crater populations produced by the OSS planetesimals characterized by the SFDs \mbox{from \cite{morbidelli2009,coradini1981}} always saturate the surface of Vesta at least to a $5\%$ level with craters that can excavate at least 7 km (the minimum thickness estimated by \cite{formisano2013,tkalcec2013}). The same is not true for the SFD \mbox{from \cite{weidenschilling2011}}: in all migration scenarios but one, the crater population excavate between 2 km and 4 km. In the scenario where Jupiter migrates by 1 au, the SFD \mbox{from \cite{weidenschilling2011}} can saturate the surface of Vesta to a $5\%$ level with craters that can excavate about 6 km. However, it must be noted that this SFD is characterized by the highest number of impact events among all the SFDs we considered, due to the high abundance of \mbox{1--2 km} wide planetesimals. Specifically, the number of impact events is high enough to cover the surface of Vesta several times with craters capable to excavate \mbox{2--4 km} of the vestan crust. Even in the migration scenario characterized by the lowest flux of OSS impactors on Vesta, \textit{i.e.}, the one where Jupiter migrates by 0.5 au, the impacts events are abundant enough to cover the surface of Vesta seven times with craters excavating about 2 km, plus about twice the vestan surface with craters that can reach down to 3--4 km. Moreover, we need to take into account the effects of the contemporary flux of ISS impactors. As we showed in Paper III, even if the bulk of the ISS impactors is characterized by lower impact velocities than OSS impactors, they can also saturate (or cover several times) the surface of Vesta with craters capable of excavating between 2 km and 4 km. When all these effects are taken into account together with the fact that impacts affected the vestan crust not only by excavating it but also by creating fractures and uncompensated negative gravity anomalies \cite{turrini2011}, it is not implausible that a fraction of the volatile materials delivered by the OSS impactors could reach the molten interior of Vesta or be incorporated into the solidifying magma brought to the surface by impact-triggered effusive phenomena.

%However, if the impacts of the OSS planetesimals can excavate the solid crust of Vesta, part of the volatile material they deliver could be trapped into the molten interior and be incorporated into the crystallizing minerals as in the samples studied by \cite{sarafian2013}. As discussed by \cite{formisano2013} and \cite{tkalcec2013}, in fact, depending on the accretion time and the initial porosity of Vesta the thickness of its solid crust across the JEB could have ranged from a minimum of $7-10$ km to a maximum of $20-30$ km. %The independent study by \cite{tkalcec2013}, using a more complete physical model, gives thickness values for the solid crust varying between $7$ km and at least $10$ km (\cite{tkalcec2013}, Supplementary Information; G. Golabek, personal communication) between $3$ Ma and $5$ Ma from CAIs, and up to $20-30$ km at about $9$ Ma (\cite{tkalcec2013}).

%%%%%%%%%%%%%%%%%%%%%%%%%%%%%%%%%%%%%%%%%%

\section{Discussion and Conclusions}

%The laboratory studies of the HED meteorites that Vesta was a planetary body mostly depleted of volatile elements at the time it solidified its basaltic crust \cite{sarafian2013}
In this work we explored the delivery of water and volatile materials to Vesta during the JEB, to assess whether the flux of OSS impactors triggered by the formation of the giant planet could be responsible for the presence of water at the time of the crystallization of some of the eucritic samples studied by \cite{sarafian2013}. In order for this to be true, one of the following two conditions needs to be fulfilled:
\begin{enumerate}
\item The water is delivered to the vestan surface and is incorporated into  eucritic magma brought on the surface by impact-triggered effusive phenomena;
\item The water is delivered directly into the molten interior of Vesta by impactors capable of excavating the solid crust of the asteroid.
\end{enumerate}

%These two conditions are not mutually exclusive and they could have both occurred on Vesta during the JEB. However, while the first condition is plausibly easier to fulfil thanks to the combined effects of the ISS and OSS impactors, the interpretation of the meteoritic data by \cite{sarafian2013} seems to favour the second condition, at least for the samples investigated to date. An additional constraint provided by the HED meteorites is that, whatever mechanism delivered the water to the asteroid, it should have influenced the eucritic magma only locally and should have preserved the global volatile-depleted nature of Vesta \cite{sarafian2013}.

These two conditions are not mutually exclusive and they could have both occurred on Vesta during the JEB. However, while the first condition is plausibly easier to fulfill thanks to the combined effects of the ISS and OSS impactors, the interpretation of the meteoritic data by \cite{sarafian2013} seems to favor the second condition, at least for the samples investigated to date. An additional constraint provided by the HED meteorites is that, whatever mechanism delivered the water to the asteroid, it should have influenced the eucritic magma only locally and should have preserved the global volatile-depleted nature of Vesta \cite{sarafian2013}.

Before we can compare the results of this study with the picture supplied by the meteoritic data, however, we need to put them into the context of the results obtained in Paper III when studying the erosive effects on Vesta of the ISS impactors caused by the JEB. According to Paper III, the SFD \mbox{from \cite{chambers2010}} is (barely) consistent with the survival of the basaltic crust of Vesta and with the observational data supplied by the Dawn mission only if Jupiter did not migrate or migrated by 0.25 au or less. However, if Jupiter did not migrate the OSS impactors would cause a large-scale impact across the JEB, with major implications for the composition of the surface of the asteroid. If instead Jupiter migrated by about 0.25 au, the flux of OSS impactors would be too low to cumulatively produce one impact: the delivery of water to Vesta would therefore be due to stochastic, low-probability events.

Also in the case of the SFD by \cite{coradini1981} the results of Paper III would indicate a limited (0.25 au) migration of Jupiter as the scenario most compatible with the observational data. The results of this study mostly confirm the compatibility of the scenario, as the erosion caused by the OSS impactors would be limited. The delivery of water would be due to a few impact events that would be capable of penetrating the solid crust even if its thickness was of the order of the highest values reported by \cite{formisano2013,tkalcec2013}. However, these impacts would create 100--200 km wide craters and could in principle bring to the surface significant quantities of diogenitic material, which is currently not observed outside the Rheasilvia basin at the south pole of Vesta \cite{desanctis2012a}. Dedicated impact simulations and geophysical studies, aiming to assess the interior state of Vesta at the time of the JEB and how it affects the outcome of the impacts, are needed before it is possible to assess whether this SFD is compatible with the observational data or not. As an example, the mechanical behavior of a solid crust of limited thickness floating on top of a molten mantle (e.g., the crust can break and/or sink more easily or, conversely, can be eroded less efficiently) can affect the delivery efficiency of water to the interior of Vesta. Alternatively, the existence of an undifferentiated primordial crust like the one suggested by the results of \cite{formisano2013,tkalcec2013} can prevent the diogenitic material from being brought to the surface even by such large impact events.

%y, can be eroded less efficiently) can affect the delivery efficiency of water to the interior of Vesta. Alternatively, the existence of an undifferentiated primordial crust like the one suggested by the results of \cite{formisano2013,tkalcec2013} can prevent the diogenitic material from being brought to the surface even by such large impact events.

According to Paper III, the most compatible SFDs are those from \cite{morbidelli2009,weidenschilling2011}. The former is compatible with the observational data on the composition of the basaltic surface of Vesta for Jupiter not migrating or migrating up to 0.25 au. However, in the scenario where Jupiter did not migrate, the SFD \mbox{from \cite{morbidelli2009}} would be associated to a significant erosion of the vestan surface and it would be compatible with the observational data only if the asteroid possessed an undifferentiated primordial crust as suggested by \cite{formisano2013}. If Jupiter did not migrate, moreover, the results of this study indicate that the flux of OSS impactors would be large enough to saturate the surface of Vesta to at least a $50\%$ level with craters 100--200 km wide and capable to excavate more than 10 km in the solid crust. Given the larger flux of OSS impactors associated to the no migration scenario, Vesta could receive enough water and volatile elements to become as enriched in water as the Earth, even when we account for the factor of two uncertainty in the water delivery discussed in Section \ref{results-water}. Unless the loss process we discussed above significantly reduced the budget of volatile elements of the asteroids, this scenario would plausibly result in a more frequent signature for the presence of water at crystallization in eucrites. If the presence of water was a local phenomenon as discussed by \cite{sarafian2013}, then also for the SFD from \cite{morbidelli2009} the most plausible scenario would appear the one where Jupiter migrated by about 0.25 au.

Finally, Paper III argued that the SFD most compatible with the observational data is the one \mbox{from \cite{weidenschilling2011}}. If Vesta already formed its basaltic crust at the time of the JEB, Paper III indicates as the most plausible scenarios those where Jupiter migrated by 0.25 au or less. If instead Vesta still possessed an undifferentiated crust as discussed by \cite{formisano2013,tkalcec2013}, Paper III favors a Jovian migration between 0.50 au and 1.00 au. In all the cases where Jupiter migrates the erosion caused by the OSS impactors is limited and the joint action of ISS and OSS impactors should allow for the solid crust to be penetrated locally to deliver water to the molten interior of the asteroid. As discussed for the previous SFD, also in the case of the SFD from \cite{weidenschilling2011} the scenario where Jupiter did not migrate could possibly be associated to too high values of the erosion of the surface and of the efficiency in the delivery of water to Vesta (again, even accounting for the factor of two uncertainty discussed in Section \ref{results-water}). In all the other scenarios, Vesta would receive enough water to be about one order of magnitude less water-rich than the Earth.

%Paper III favours a Jovian migration between 0.50 au and 1.00 au. In all the cases where Jupiter migrates the erosion caused by the OSS impactors is limited and the joint action of ISS and OSS impactors should allow for the solid crust to be penetrated locally to deliver water to the molten interior of the asteroid. As discussed for the previous SFD, also in the case of the SFD from \cite{weidenschilling2011} the scenario where Jupiter did not migrate could possibly be associated to too high values of the erosion of the surface and of the efficiency in the delivery of water to Vesta (again, even accounting for the factor of two uncertainty discussed in Section \ref{results-water}). In all the other scenarios, Vesta would receive enough water to be about one order of magnitude less water-rich than the Earth.

%The meteoritic data presently available on the interior of Vesta seem to indicate that the water delivery process in the asteroid belt was efficient (\textit{i.e.}, it needed to deliver to Vesta enough water to allow for a reasonable chance of implanting some of it into its molten interior) but not too efficient (\textit{i.e.}, Vesta should retain its volatile-depleted mineralogical signature). This would translate, according to our results, in the requirement that Jupiter migrated by a significant fraction of au. Once we take into account the global distribution of the material from the impactors on the vestan surface as shown in Figures \ref{fig-crater} and \ref{fig-sprea

The meteoritic data presently available on the interior of Vesta seem to indicate that the water delivery process in the asteroid belt was efficient (\textit{i.e.}, it needed to deliver to Vesta enough water to allow for a reasonable chance of implanting some of it into its molten interior) but not too efficient (\textit{i.e.}, Vesta should retain its volatile-depleted mineralogical signature). This would translate, according to our results, in the requirement that Jupiter migrated by a significant fraction of au. Once we take into account the global distribution of the material from the impactors on the vestan surface as shown in Figures \ref{fig-crater} and \ref{fig-spreading}, our results indicate that the surface of the asteroid could have contained significant quantities of exogenous water and volatile materials mixed with the regolith at the time of the JEB in all considered migration scenarios. Moreover, the results of Paper I indicate that close encounters with the forming core of Jupiter could cause OSS planetesimals to impact Vesta even before the giant planet started accreting its gaseous envelope (albeit with lower frequency than during the JEB), strengthening the link between the appearance of volatile materials on Vesta and the formation of the giant planet. The data supplied by the Dawn mission \cite{desanctis2012b,prettyman2012,denevi2012,mccord2012} indicate that hydrated materials have been delivered in significant quantities to Vesta over the last 4 Ga and presently affect about 30$\%$ of the surface of the asteroid \cite{mccord2012,prettyman2012,turrini20XX}. As we said, we don't expect the hydrated and H-rich material presently found of Vesta to date back to the JEB, as the more abundant ISS impactors would plausibly remove the vast majority of the water and volatile elements that did not reach the molten interior of the asteroid. However, it is interesting to note that these results globally suggest that the presence of volatile materials and, likely, of pitted terrains on the surface of Vesta could have been a continuous phenomenon during the life of the asteroid.

We must take into account, however, that the results we present in this work can be significantly affected by the following sources of uncertainty. The first source of uncertainty is linked to the unknown composition of the primordial planetesimals that impacted Vesta, which as we said before should introduce at least a factor of 2 uncertainty in the estimated amounts of water delivered to the asteroid. The second source of uncertainty is due instead to the distribution of the material from the impactors on the surface of the asteroid, which depends on the unknown composition and porosity of the impacting bodies and on their impact angle, and which affects the quantity of water that can penetrate the crust. Another important source of uncertainty in the interpretation of our results is the thickness of the solid crust of Vesta at the time of the JEB. The range of values reported by \cite{formisano2013,tkalcec2013} spans \mbox{between 7 and 30 km}: if the JEB took place when the solid layer was toward the high-end tail of these values, more OSS impactors would be required to have a reasonable chance of a few impactors reaching the molten interior of Vesta. Moreover, the hypothesized presence of an undifferentiated crust would favor more erosive scenarios, and therefore higher fluxes of impactors, than those requested to preserve an already formed basaltic crust. The presence of a thick solid layer would therefore favor the scenario where Jupiter did not migrate, while the existence of an undifferentiated crust would probably require larger displacements of Jupiter (between 0.50 au and 1 au) to allow for its removal. Finally, the dynamical model on which the simulations of Paper I were based did not account for the effects of the perturbations of planetary embryos and of gas drag. These two competing effects act respectively to rise and to damp the orbital eccentricities and inclinations of the planetesimals and therefore affect both the intensity of the fluxes of impactors and the impact velocities (see Papers I, II and III for further discussions).

% favour more erosive scenarios, and therefore higher fluxes of impactors, than those requested to preserve an already formed basaltic crust. The presence of a thick solid layer would therefore favour the scenario where Jupiter did not migrate, while the existence of an undifferentiated crust would probably require larger displacements of Jupiter (between 0.50 au and 1 au) to allow for its removal. Finally, the dynamical model on which the simulations of Paper I were based did not account for the effects of the perturbations of planetary embryos and of gas drag. These two competing effects act respectively to rise and to damp the orbital eccentricities and inclinations of the planetesimals and therefore affect both the intensity of the fluxes of impactors and the impact velocities (see Papers I, II and III for further discussions).

Addressing these sources of uncertainties, to shed light on the details and the implications of the flux of cometary objects in the asteroid belt caused by the formation of Jupiter, will require further dedicated studies. Notwithstanding these uncertainties, however, the qualitative picture described by the results of this work appears robust, as the formation of Jupiter can deliver water and volatile elements to Vesta at a time when the thickness of the vestan crust was limited and part of the eucritic layer was still in a molten state as required to explain the samples studied by \cite{sarafian2013}. More generally, the results here presented highlight the fact that the Jovian Early Bombardment could represent a viable mechanism for the delivery of water into the asteroid belt and possibly up to the orbital region of Mars \cite{weidenschilling1975}. Such mechanism would be more efficient across the asteroid belt (particularly beyond about 2.5 au, see Papers I and II) than at Mars \cite{weidenschilling1975}, but the results of theoretical studies indicate that part of the planetary material residing between Mars and the outer bound of the asteroid belt could be later accreted by the forming terrestrial planets (see \cite{righter2011,morbidelli2012} and references therein). This two-phases mechanism for delivering water to the forming terrestrial planets would be in agreement with the global picture supplied by the results of \cite{hartogh2011}, which attributes an important role to comets in the delivery of water to the inner Solar System. The investigation of the efficiency of this mechanism in the Solar System, in turn, could help us improve our understanding of the formation and early evolution of Jupiter and, therefore, of the Solar Nebula. Finally, as pointed out by Paper II the Jovian Early Bombardment is due to physical processes that are general to planetary systems harbouring forming giant planets. The Jovian Early Bombardment therefore represents a specific case of a more general class of events, the Primordial Heavy Bombardments \cite{coradini2011,turrini2012}, and the results of this work can be applied also to the study of extrasolar planetary systems. %These phases of intense bombardment and remixing of the solid materials triggered by the formation of giant planets in the protoplanetary disk in which they are embedded

%In order to compare the results of this study with the picture supplied by the meteoritic data, however, we first need to put them into the context the geophysical state of Vesta at the time of the JEB. In particular, we need to take into consideration the thickness of the solidified crust of the asteroid. If the thickness of the solid crust was similar to the higher values reported by \cite{formisano2013,tkacec2013}
%If the solid crust was characterized by a more or less uniform thickness, then 

%%%%%%%%%%%%%%%%%%%%%%%%%%%%%%%%%%%%%%%%%%

\section*{\noindent Acknowledgments}
\vspace{12pt}

The authors would like to thank Stuart Weidenschilling, Guy Consolmagno, and Maria Teresa Capria for their assistance and suggestions in defining the initial conditions and the observational constraints for this work, and the two anonymous reviewers whose comments helped improve the quality of this work. This research has been supported by the Italian Space Agency (ASI) through the ASI-INAF contract I/010/10/0, by the Russian Fund of Basic Research through Grant 13-05 -00694-a, and by the International Space Science Institute in Bern through the International Teams 2012 project ``Vesta, \mbox{the key to} the origins of the Solar System'' \cite{issi2012}. The computational resources used in this research have been supplied by INAF-IAPS through the projects ``HPP---High Performance Planetology'' and ``DataWell''.

\section*{\noindent Conflicts of Interest}
\vspace{12pt}

{The authors declare no conflict of interest.}

%=================================================================
% References: Variant A
%=================================================================
% Back Matter (References and Notes)
%----------------------------------------------------------
% Style and layout of the references
\bibliographystyle{mdpi}
\makeatletter
\renewcommand\@biblabel[1]{#1. }
\makeatother

%=================================================================
% References:  Variant B
%=================================================================
% Use the following option to include external BibTeX files:
%\bibliography{lite}
%\bibliographystyle{mdpi}

\end{document}